\documentclass[aps,prb,reprint,superscriptaddress,amsmath,amsfonts,amssymb]{revtex4-2}

\usepackage{hyperref}
\usepackage{physics}
\usepackage{xstring}
\usepackage{float}
\usepackage{graphicx}
\usepackage{dcolumn}
\usepackage{bm}
\usepackage{amsmath}
\usepackage{amssymb}
\usepackage{xcolor}
\usepackage[normalem]{ulem}

\usepackage{cancel}
\usepackage{multirow}
\usepackage[abs]{overpic}
\usepackage{array}
\usepackage{nicefrac}

\usepackage{lineno}
\makeatletter
\patchcmd{\frontmatter@abstract@produce}
{\vskip200\p@}
{\linenumbers\vskip200\p@}
{}{}
\makeatother
\newcommand{\Tord}[1]{\mathrm{T}\left\{#1\right\}}

\newcommand{\uba}{Universidad de Buenos Aires, Facultad de Ciencias Exactas y Naturales, Departamento de F\'\i sica, Buenos Aires, Argentina}
\newcommand{\ifiba}{CONICET-Universidad de Buenos Aires, Instituto de F\'\i sica de Buenos Aires, Buenos Aires, Argentina}

\begin{document}
\title{Quantum Thermodynamic Uncertainty Relation and Macroscopic Superconducting  Coherence}
\author{Franco Mayo}
\affiliation{\uba} \affiliation{\ifiba} \affiliation{The Abdus Salam International Center for Theoretical Physics, Strada Costiera 11, 34151 Trieste, Italy}
\author{Nahual Sobrino}\email{nsobrino@ictp.it}
\affiliation{The Abdus Salam International Center for Theoretical Physics, Strada Costiera 11, 34151 Trieste, Italy}
\author{Rosario Fazio} 
\affiliation{The Abdus Salam International Center for Theoretical Physics, Strada Costiera 11, 34151 Trieste, Italy}
\affiliation{Dipartimento di Fisica, Universit\`a di Napoli ``Federico II”, Monte S. Angelo, I-80126 Napoli, Italy}
\author{Fabio Taddei}
\affiliation{NEST, Istituto Nanoscienze-CNR and Scuola Normale Superiore, Piazza San Silvestro 12, I-56127 Pisa, Italy}
\author{Michele Governale}
\affiliation{School of Chemical and Physical Sciences and MacDiarmid Institute for Advanced Materials and Nanotechnology,
Victoria University of Wellington, PO Box 600, Wellington 6140, New Zealand}

\begin{abstract}
Stability and efficiency are mutually exclusive in a thermodynamic process, e.g. in a thermal machine. Any effort to reduce the fluctuations of a certain output quantity is necessarily accompanied by an increase of entropy production, therefore lowering its efficiency. This interplay is beautifully captured by the so called Thermodynamic Uncertainty Relations (TURs) which set a lower bound on the relative uncertainty of a current for a given rate of entropy production. Their status in hybrid normal--superconducting (N--S) devices has remained unsettled. We show that, in the subgap regime, departures from the normal quantum TUR are governed by {\it macroscopic} superconducting coherence  quantified by the pair amplitude, and that introducing a dephasing probe suppresses this coherence and restores the bound. We further derive a hybrid quantum TUR that is general for two-terminal N--S junctions in the Andreev regime: the inequality is never violated, is saturated only at vanishing current, and is related to the normal quantum bound under the replacement \(e\to 2e\). For N--S quantum dot and Cooper-pair-splitter systems we compute current and noise and show that deviations from the normal bound track the pair amplitude on the central region. The results establish a direct link between superconducting	macroscopic coherence and nonequilibrium fluctuations and supply a general bound for
the Andreev regime.
\end{abstract}

\maketitle

\textit{Introduction---}
The TUR for classical systems governed by Markovian dynamics can be expressed as the inequality $\sigma \Delta I/(k_BI^2)\geq 2$~\cite{barato2015thermodynamic, gingrich2016dissipation, pietzonka2016universal, timpanaro2019thermodynamic}, where $I$ is the  current associated to a thermodynamic process, $\Delta I$ its variance, $\sigma$ the entropy production rate and $k_B$ the Boltzmann constant. The study of this and related bounds has attracted recently a lot of attention to understand under which circumstances they can be violated, and consequently,  to search for inequalities that account for different dynamical regimes. An account of this activity can be found in Refs.~\cite{Seifert2019,horowitz2020thermodynamic}. More specifically, TURs in electronic-transport setups have been considered in several papers both in the static~\cite{agarwalla2018assessing, Ehrlich2021,  gerry2022absence,  Guarnieri2019, Kheradsoud2019, Proesmans2019,liu2019thermodynamic, Pietzonka2018, Prech2023, Ptaszynski2018, Saryal2019, Saryal2021, Timpanaro2025, Brandner2018, macieszczak2018unified, Saryal2022, gingrich2016dissipation, braggio2011superconducting, taddei2023thermodynamic, horowitz2020thermodynamic, Kamijima2021, Ohnmacht2024, Lopez2023, Manzano2023Oct, Misaki2021, Palmqvist2024, Wozny2025,Zhang2025,Brandner2025} and time-dependent case~\cite{Potanina2021,Lu2022}.

For charge currents, it is convenient to re-express the TUR employing the Fano factor $F=\mathcal{S}/ (e |J|)$, in the form 
\begin{equation}
\frac{F}{\mathcal{B}\left[2 k_B |J|/(e\sigma)\right]}-1\geq 0 ,
\label{eq:ine}
\end{equation}
where the function $\mathcal{B}[x]$ depends only the dimensionless quantity $x=2 k_B |J|/(e\sigma)$, with $J$ being the charge current, $\mathcal{S}$ the noise, $\sigma$ the entropy production rate, $k_B$ the Boltzmann constant and $e>0$ is the magnitude of the electron charge.
By simply imposing the positivity of the entropy production rate in the linear response regime, the function $\mathcal{B}$ is found to take the form~\cite{barato2015thermodynamic,macieszczak2018unified,Proesmans2019,taddei2023thermodynamic}
$\mathcal{B_{\rm cl}}(x)=x$, which is equivalent to the TUR reported above and originally derived for classical Markovian processes.
The inequality~\eqref{eq:ine} with the function $\mathcal{B_{\rm cl}}$, that we will refer to as the {\it classical TUR}, is generally found to hold, but fails in the presence of quantum coherence.
This is the case of quantum dot systems at strong coupling, i.e. when (non-classical) high-order tunneling processes contribute to transport~\cite{agarwalla2018assessing,Ptaszynski2018,Brandner2018,liu2019thermodynamic, Ehrlich2021, gerry2022absence, Prech2023, Manzano2023Oct,Timpanaro2025}.
Such violations of the classical TUR are a tell-tale for the quantum-coherent nature of transport.
A step forward, to include quantum coherent effects in determining a new bound, has been recently achieved in Ref.~\cite{Brandner2025}. 
A different inequality, the {\it quantum TUR}, has been derived making use of the Landauer-B\"uttiker scattering theory of quantum transport.  In this case, the function $\mathcal{B}$ is given by $\mathcal{B_{\rm qu}}\left( x \right)= \text{cosech}\left(1/x\right)$ and is never violated for non-interacting electrons in a quantum coherent conductor.

An interesting question now arises if the presence of {\em macroscopic coherence} adds new features to the picture, i.e. whether the quantum TUR should be obeyed in superconducting systems, where coherence has a macroscopic origin related to the superconducting condensate of Cooper pairs. TURs in hybrid superconducting systems have been considered in a few papers only~\cite{Misaki2021,Manzano2023Oct,Lopez2023,taddei2023thermodynamic,Ohnmacht2024}. Clear violations of the classical TUR have been demonstrated in such systems~\cite{Manzano2023Oct,Lopez2023,Ohnmacht2024}, where Andreev reflection plays a key role in exceeding this bound.  What happens to the quantum inequality? Equivalently, how is the interplay between fluctuations and entropy production affected by the presence of macroscopic coherence?

The aim of this work is to answer this question by studying hybrid superconductor-normal systems and providing a new general bound for these class of systems. 

\begin{figure}[!ht]
\includegraphics[width=0.48\textwidth]{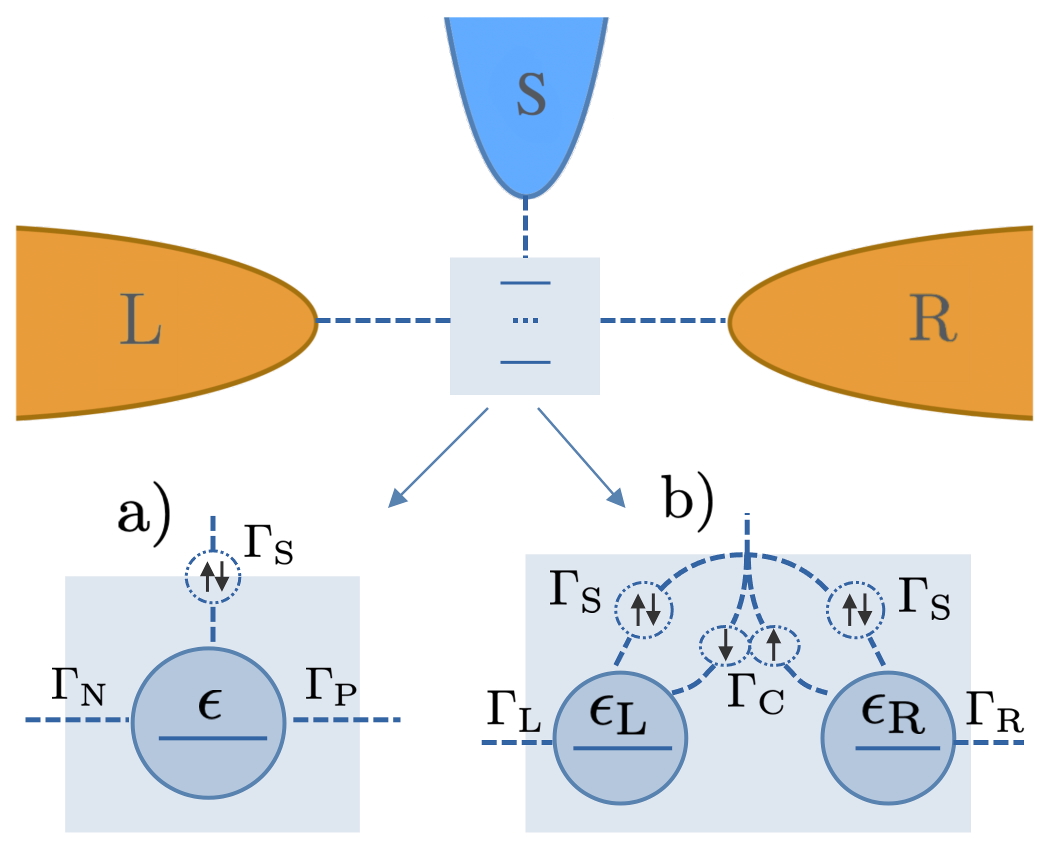}
\caption{ A central region is coupled to a superconducting (S) lead with tunneling rate $\Gamma_{\rm S}$, and to two  normal leads $\eta={\rm L,R}$ with tunneling rates $\Gamma_{\rm \eta}$, chemical potential $\mu_{\rm \eta}$ and kept at temperature $T$. a) The central region corresponds to a quantum dot (with energy level $\epsilon$) and one of the normal leads is considered as a probe coupled to the dot by a tunneling rate $\Gamma_{\rm P}$ and with a chemical potential $\mu_{\rm P}$ that is adjusted so that the current flowing in P is zero. b) The central region is a double quantum dot. The superconducting lead is also coupled to the dots through the nonlocal tunneling rate $\Gamma_{\mathrm{C}}$.}
\label{fig:diagram}
\end{figure}

While single particle coherence does not lead to any violation of Eq.(\ref{eq:ine}) (in both cases)
macroscopic coherence does.
We will consider a system consisting of a superconducting lead and two normal leads coupled to a region containing localized levels, as shown in Fig.~\ref{fig:diagram}.
By using a Green's functions approach, we look for violations of both forms of TUR in the case where the central region is either a single (Fig.~\ref{fig:diagram}a) or two (Fig.~\ref{fig:diagram}b) QDs, to see also the effect of non-locality introduced by the crossed Andreev reflection.
Dephasing is introduced by  using lead R as a voltage probe. In this case the violation of the quantum TUR weakens by increasing the coupling to the probe, i.e.,~by suppressing the superconducting coherence, as measured by the superconducting pair amplitude of the QD.
Our analysis is further supported by the derivation of a new inequality that holds  for two-terminal normal–superconducting junctions in the $|\Delta|\!\to\!\infty$ limit. This limit is commonly adopted since it effectively captures the essential physics of subgap transport dominated by coherent Andreev processes, and is equivalent to assuming temperatures and applied voltages much smaller than the superconducting gap \cite{martin2011josephson}.

\textit{Model and Formalism---} Let us consider a central region containing localized levels (QDs), coupled to one superconducting (S) lead and two normal leads $\eta={\rm L,R}$, as shown in Fig.~\ref{fig:diagram}.
The normal leads are described by the Hamiltonians
$ H_\eta =\sum_{k, \sigma}\epsilon_k\, c^\dagger_{\eta k\sigma}\,c_{\eta k\sigma}$,
where $c_{\eta k \sigma}$ is the annihilator operator for an electron with momentum $k$ and spin $\sigma$ in lead $\eta$. The single-particle energies in the leads are denoted by $\epsilon_k$ and we assume that are the same for both leads and independent of spin.
The leads are coupled to the central region by means of  the standard tunneling Hamiltonians: 
$H_{{\rm tunn},\eta} =\sum_{k,\sigma,i}  V_{\eta,i}\left(c^\dagger_{\eta k\sigma}d_{i,\sigma} + \text{H.c.}\right)$,
where the label $i$ denotes a localized state in the central region and will be omitted for the case of only one quantum dot, $d_{i\sigma}$ is the annihilation operator for an electron in level $i$, and without loss of generality, we take $V_{\eta,i}$ to be real.
Both normal leads are kept at temperature $T$ with chemical potentials $\mu_{\rm L}$ and $\mu_{\rm R}$, respectively. 
The superconducting lead is grounded and we assume the gap $\Delta$ of the superconductor  to be the largest energy scale ($|\Delta|\rightarrow\infty$ limit). In this limit, the quasi-particles in the superconductor do not contribute and the Andreev reflection processes  between the superconductor and the central region can be included as an effective pairing term in the Hamiltonian of the central region~\cite{rozhkov2000interacting, meng2009self, eldridge2010superconducting}.

The current $J_\eta$ flowing from lead $\eta$ into the central region can be expressed in terms of the full Green's functions of the central region in Nambu representation and the Fermi function of lead $\eta$ \cite{pala2007nonequilibrium}. 
The other ingredient needed for the TUR is the zero-frequency noise~\cite{beenakker1994,anantram1996current,cao2015currents} in lead S, which is defined as $\mathcal{S} = \frac{1}{2}\,\int dt\, \langle J_{\rm S} (t)J_{\rm S}(0) + J_{\rm S}(0)J_{\rm S}(t)\rangle - \langle J_{\rm S}\rangle \langle J_{\rm S}\rangle$.
If the interactions in the central region can be treated at the level of mean-field, also the zero-frequency noise can be written in terms of Green's function of the central region in Nambu representation and the Fermi function of lead $\eta$. The expressions for the current, the noise and the Green's functions of the central region can be found in the Supplemental Material~\cite{SM}. 
Since the normal currents $J_{\eta}$ are the only dissipative currents in the system, the entropy-production rate is simply given by $\sigma=- \sum_{\eta}\mu_{\eta} J_{\eta}/(eT)$, where $-\mu_{\eta}/e$ is the voltage applied to each normal lead.

\textit{Effect of Macroscopic coherence on TUR---}The effects of macroscopic quantum coherence can be explored by modeling the central region in its minimal form as a single QD. 
The right normal lead will be used as a dephasing probe (R=P) by setting its chemical potential $\mu_{\rm R}=\mu_{\rm P}$ in such a way to nullify the current flowing through it. The left normal lead (L=N) is kept at a chemical potential $\mu_{\rm L}=\mu_{\rm N}$. We denote the tunneling rates between the leads P and N and the QD,  as $\Gamma_P$ and $\Gamma_N$, respectively. These rates are calculated in second-order in $H_{{\rm tunn},\eta}$ in the wide-band limit.

In the $|\Delta|\rightarrow\infty$ limit,
one can write an effective Hamiltonian for the QD that takes into account the coupling to the superconductor~\cite{rozhkov2000interacting, meng2009self, eldridge2010superconducting}: 
\begin{equation}
H_{\rm QD} = \epsilon \sum_\sigma d^\dagger_\sigma d_\sigma - \frac{\Gamma_{\rm S}}{2}\,\left(d^\dagger_\uparrow d^\dagger_\downarrow + \,d_\downarrow d_\uparrow \right)\:.
\nonumber
\end{equation}
Here, $\epsilon$ represents the energy level of the QD.
We start with the ideal system, where decoherence is absent, by setting $\Gamma_{\rm P}=0$. 

The exact expressions for the current $J$ flowing from the lead N into the QD and its zero-frequency noise are provided in the Supplemental Material~\cite{SM}. 

Here, we  examine the limit where $\Gamma_{\rm N}\ll k_B T$, but making no assumptions on $\epsilon$, $\Gamma_{\rm S}$ and $\mu_{\rm N}$. 
This is in contrast to standard perturbative expansions in $\Gamma_N$, as e.g. in \cite{braggio2011superconducting}, where it is also assumed that $\Gamma_N$ is much smaller than the splitting of the Andreev bound states, $\epsilon_A=\sqrt{\epsilon^2+\frac{\Gamma_{\rm S}^2}{4}}$, and therefore of $\Gamma_S$. The standard approach leads to expressions that are not valid in the region of parameter space where TUR violations occur.
In the case considered here, the
current and noise can be written as 
\begin{align}
\label{JS1}
J = & -\frac{e}{h}\, \pi\frac{\Gamma_{\rm N} \frac{\Gamma_{\rm S}^2}{2}}{  \bar{\Omega}^2}
\Tr \left[\boldsymbol{\tau_3 F^+_{\rm N}}(\epsilon_A) \right], \\
\mathcal{S}=& \frac{e^2}{h} \,  \pi \frac{\Gamma_{\rm N} \frac{\Gamma_{\rm S}^2}{2}}{\bar{\Omega}^3 }\left\{\left( \bar{\Omega}^2 + \Gamma_{\rm N}^2\right)\frac{\Gamma_{\rm S}^2}{4} 
\Tr\left[\boldsymbol{F^+_{\rm N}}(\epsilon_A)\boldsymbol{F^-_{\rm N}}
(-\epsilon_A)\right]\nonumber \right.
\\
& +  \left.
\left[
2 \bar{\Omega}^4 - \left(\bar{\Omega}^2 + \Gamma_{\rm N}^2\right)\frac{\Gamma_{\rm S}^2}{4}\right]
\Tr\left[\boldsymbol{F^+_{\rm N}}(\epsilon_A)\boldsymbol{F^+_{\rm N}}(-\epsilon_A)\right]
\right\},
\label{JS2}
\end{align}
where $\bar{\Omega}^2=\epsilon_A^2+\Gamma_{\rm N}^2/4$ and $\boldsymbol{\tau_3}$ is the $z$ Pauli matrix. 
The symbol $\boldsymbol{F^{\pm}_{\eta}(\omega)}$ denotes the $2\times 2$ diagonal  matrices 
$\boldsymbol{F^{\pm}_{\eta}(\omega)} = \text{Diag}[f^\pm(\omega - \mu_\eta), f^\mp(-\omega- \mu_\eta)]$,
where $f^+(\omega)\equiv f(\omega)$ and $f^-(\omega) \equiv 1-f(\omega)$ with  $f(\omega)$ being the Fermi function at temperature $T$.
In this limit, the Fano factor is minimized for 
$ \epsilon=0$  and $\Gamma_{\rm S}=\sqrt{5/3}\,\Gamma_{\rm N}$,
taking a very simple form
\begin{align}
F=&\frac{9}{16}\left[1-2 f(-\mu_{\rm N})\right]-\frac{1}{1-2 f(-\mu_{\rm N})},
\end{align}
when $\Gamma_{\rm N}\ll \mu_{\rm N}$ is further assumed.
This expression breaks the quantum and classical bounds when 
\begin{align*}
&{|\mu_{\rm N}|}/({k_B T})< 2 \log\left[1/7 (9 + 4 \sqrt{2})\right]\approx 1.478\quad&(F<\mathcal{B_{\rm qu}})\\
&{|\mu_{\rm N}|}/({k_B T})\lesssim 4.074&(F<\mathcal{B_{\rm cl}}).
\end{align*}
This shows that both the classical and quantum-coherent TUR can be violated for small values of $\Gamma_{\rm N}$  as long as $\Gamma_S \approx \Gamma_N$.
This contrasts with what happens in the corresponding normal system, where the S lead is in the normal phase. In this case, violations of the classical TUR occur only at strong coupling, i.e.~when (non-classical) high-order tunneling processes contribute to transport~\cite{agarwalla2018assessing,Ptaszynski2018}.

\begin{figure}[ht!]
\centering
\includegraphics[width=0.5\textwidth]{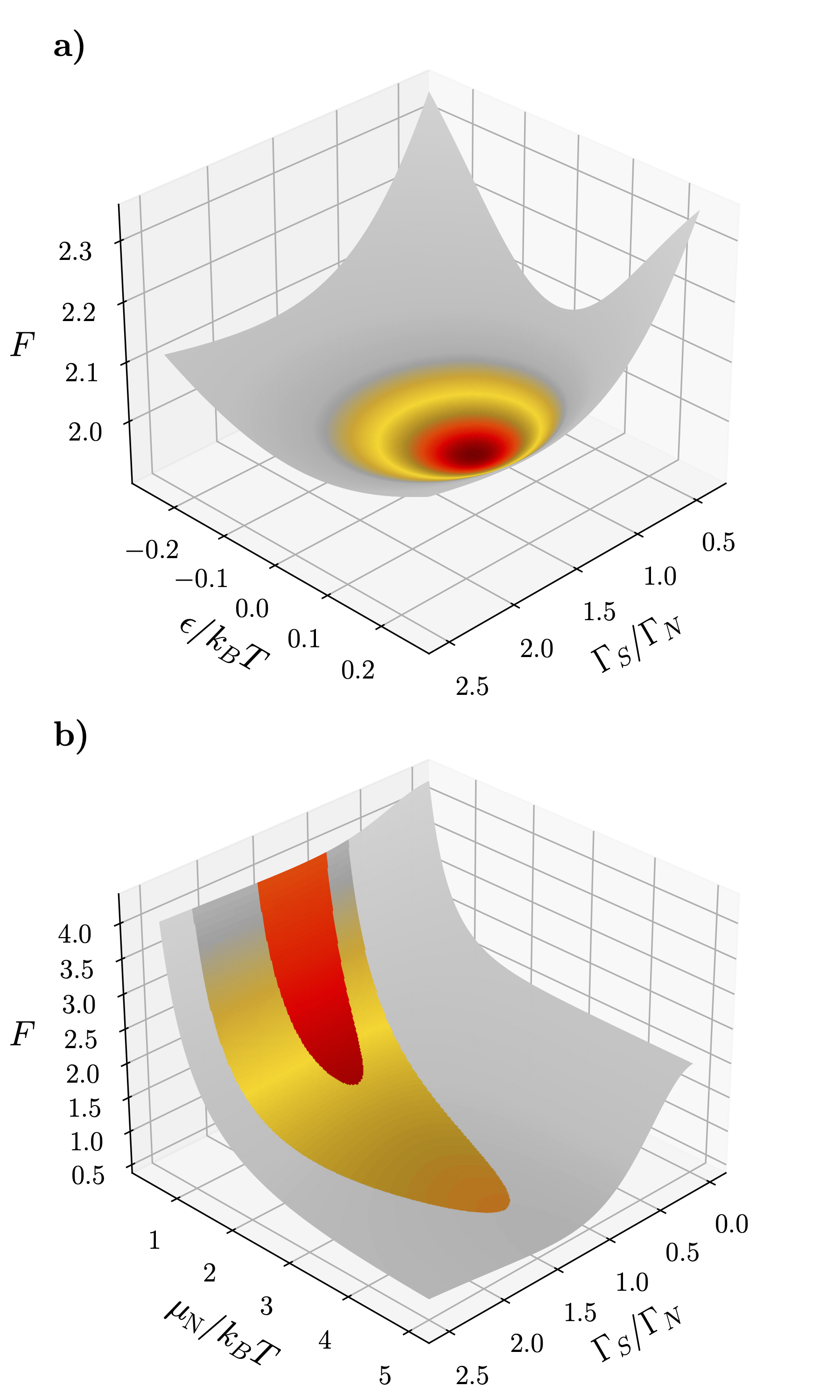}
\caption{Fano factor as a function of (a) $\epsilon$ and (b) $\mu_{\rm N}$ and the ratio $\Gamma_{\rm S}/\Gamma_{\rm N}$ . The red area shows the region in which the quantum bound $\mathcal{B_{\rm qu}}$ is broken, and the yellow area corresponds to violations of the classical bound $\mathcal{B_{\rm cl}}$. Parameters (when fixed): $\Gamma_{\rm N} = 0.6 k_BT$, $\mu_{\rm N} = k_BT$ and $\epsilon = 0$. 
}
\label{fig:results1}
\end{figure}
\begin{figure}[!ht]

\includegraphics[width=0.45\textwidth]{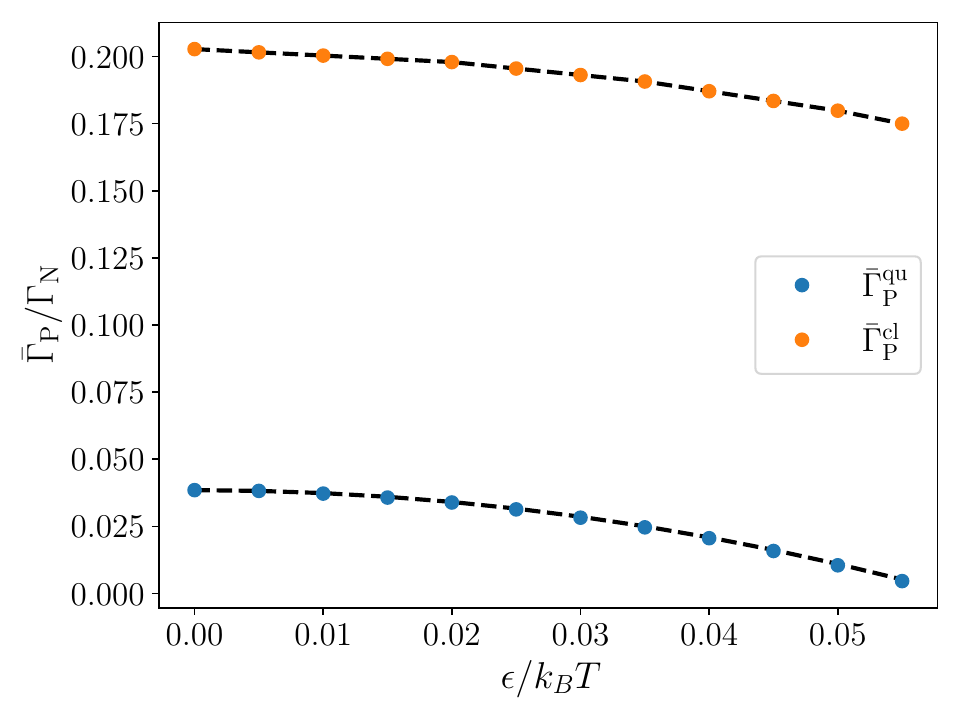}
\caption{The value of $\Gamma_{\rm P}$ for which the decoherence is strong enough that the quantum (blue) and classical (orange) TUR stop being violated as a function of $\epsilon$.
	Parameters: $\Gamma_{\rm N}=0.6k_B T$, $\Gamma_{\rm S} = \sqrt{\frac{5}{3}}\Gamma_{\rm N}$ and $\mu_{\rm N}=k_B T$.}
\label{fig:probe}
\end{figure}

To go beyond the small coupling regime, in
Fig.~\ref{fig:results1} we plot the Fano factor for  $\Gamma_{\rm N}=0.6k_BT$, obtained by numerical integration of the full expressions for $J$ and $\mathcal{S}$ provided in Supplemental Material~\cite{SM}.
In panel (a), $F$ is plotted as a function of the level position and the tunnel-coupling ratio $\Gamma_{\rm S}/ \Gamma_{\rm N}$ for fixed $\mu_{\rm N}=k_BT$, while in panel (b) $F$ is plotted as a function of chemical potential and the tunnel-coupling ratio for fixed $\epsilon=0$.
The region of parameter space where both bounds are violated are colored in red, while in the yellow region only the classical bound is violated.
According to panel (a), also in this situation we find that both bounds are violated in a region of parameter space around $\epsilon=0$ and similar coupling to the leads S and N.
More interestingly, there is also a smaller region (red part of the surface) where the quantum bound $\mathcal{B_{\rm qu}}$ is also violated.
We also find that there is a maximum value for the magnitude of the chemical potential to observe TUR violations [see paned (b) of Fig~\ref{fig:results1}]. This is trivially due to the fact that for large $\mu_N$ the entropy-production rate suppresses both bounds $\mathcal{B}_{\rm cl}$ and $\mathcal{B}_{\rm qu}$.
Note that $F$ is even in $\mu_{\rm N}$ and diverges as $\mu_{\rm N}$ goes to zero since the current $J$ vanishes.
We note that the Fano factor obtained from Eqs.~\eqref{JS1} and \eqref{JS2} agrees quantitatively very well with the numerical results for $\Gamma_{\rm N}\le 0.1 k_B T$. For larger values of $\Gamma_{\rm N}$, as the one in Fig.~\ref{fig:results1}, we still find a good qualitative agreement.

Physically, the violation of the quantum bound can be attributed to the presence of macroscopic quantum coherence due to the presence of the superconducting condensate. The superconducting correlations in the QD can be quantified by the pair amplitude 
$|\langle d_{\downarrow}d_{\uparrow}\rangle|=\left| \int \frac{d\omega}{2\pi} 
\left(\mathbf{G}^<(\omega)\right)_{1,2}\right|$.  
From a mathematical point of view, the violation of the quantum bound in our system is not unexpected, as it was originally derived based on currents and noise expressions for a normal system.
In contrast, in our case the expressions for current and noise have a  different structure due to the coupling between electrons and holes induced by superconducting correlations.

To corroborate the fact that the violation of the quantum bound is due to the presence of macroscopic quantum coherence, we introduce a source of decoherence by adding a fictitious voltage probe, which reduces the presence of superconducting correlations as measured by the pair amplitude. The probe consists of a normal lead (denoted by P in Fig.~\ref{fig:diagram}) whose voltage is adjusted to make the current flowing through the lead vanish. To do so we calculate the current in the probe $J_{\rm P}$ numerically for different values of $\mu_{\rm P}$ and find the one that nullifies the current. The coupling strength of the probe is set by its tunneling rate $\Gamma_{\rm P}$. We find that for small values of the probe coupling $\Gamma_{\rm P} \ll \Gamma_{\rm S}$ the pairing parameter is reduced by the noise as 
\begin{equation*}
\vert \langle d_\downarrow d_\uparrow \rangle\vert = \vert \langle d_\downarrow d_\uparrow \rangle\vert \eval_{\Gamma_{\rm P} = 0} - \alpha\frac{\Gamma_{\rm P}}{\Gamma_{\rm S}},
\end{equation*}
where $\alpha > 0$ is a parameter that depends on $\Gamma_{\rm N}$, $\epsilon$ and $\mu_{\rm N}$.
For the parameters values used for the Figures, we have found $\alpha\approx 0.05$.
As the probe introduces noise in the system we find that both the quantum and classical TUR stop being violated. We define $\bar{\Gamma}_{\rm P}^{\rm qu}$($\bar{\Gamma}_{\rm P}^{\rm cl}$) the value of the probe's coupling for which decoherence is strong enough that the quantum (classical) bound is no longer broken. In Fig.~\ref{fig:probe} we show the values of $\bar{\Gamma}_{\rm P}$ for both bounds. As expected, the presence of the noise makes both TUR violations disappear. We can see that the quantum bound is more sensible to this noise since the values of $\bar{\Gamma}_{\rm P}^{\rm qu}$ are always smaller than those of $\bar{\Gamma}_{\rm P}^{\rm cl}$, which makes sense considering that the violation of the quantum bound is always smaller than that of the classical one.

\begin{figure}[!ht]
\includegraphics[width=0.48\textwidth]{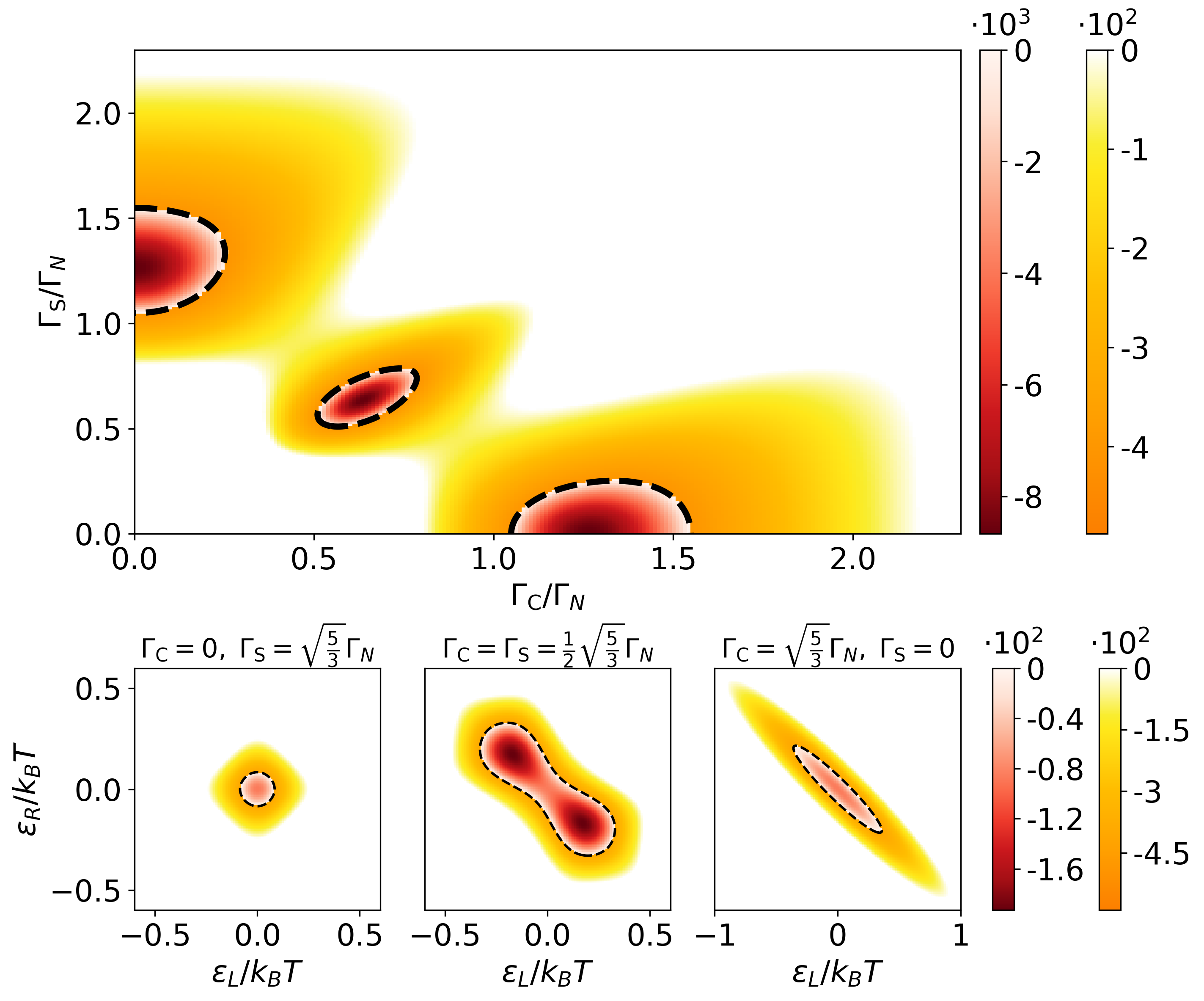}
\caption{Regions of violation of the classical (yellow) and quantum (red) TURs in the CPS system. The plotted quantities correspond to the left-hand side of Eq.~\eqref{eq:ine} where the respective inequality is violated. The top panel displays the dependence on the tunneling rates $\Gamma_{\mathrm{S}}/\Gamma_{\mathrm{N}}$ and $\Gamma_{\mathrm{C}}/\Gamma_{\mathrm{N}}$, while the bottom panels show the dependence on the energy levels $\varepsilon_{\eta}$ for the three representative points in the top panel where the violations are maximal. Simultaneous LAR and CAR processes result in the maximal violation of the TURs. The black dashed line marks the saturation of the quantum bound. Parameters (when fixed): $\Gamma_{\mathrm{N}} = 0.6\, k_{\mathrm{B}}T$, $\varepsilon_{\eta} = 0$, and $\mu_{\mathrm{N}} = k_{\mathrm{B}}T$.}
\label{fig:CPS}
\end{figure}
\textit{Nonlocal correlation effects on the TUR---}
In order to investigate the effect of non-local correlations on the TUR, we model the central region as a Cooper-pair splitter (CPS), as shown in Fig.~\ref{fig:diagram}(b).
The system consists of a double quantum dot symmetrically coupled to two normal leads ($\Gamma_{\rm L}=\Gamma_{\rm R}\equiv \Gamma_{\rm N}$) and one superconducting lead. In the $|\Delta|\rightarrow\infty$  limit, the effective Hamiltonian takes the form
\begin{align}
H_{\rm CPS} =& \sum_{\eta,\sigma}  \epsilon_{\eta} d^\dagger_{\eta,\sigma} d_{\eta,\sigma} - \frac{\Gamma_{\rm S}}{2}\sum_{\eta}\,\left(d^\dagger_{\eta,\uparrow} d^\dagger_{\eta,\downarrow} + {\rm H.c.} \right)\nonumber \\
& - \frac{\Gamma_{\rm C}}{2}\,\left(d^\dagger_{R,\uparrow} d^\dagger_{L,\downarrow} -d^\dagger_{R,\downarrow} d^\dagger_{L,\uparrow} + {\rm H.c.} \right) \: ,
\nonumber
\end{align}
where $\eta={\rm L,R}$ refers to the dots index. The last term describes crossed Andreev reflection, in which a Cooper pair is coherently split between the two dots, generating non-local superconducting correlations characterized by the rate $\Gamma_{\rm C}$.

The combined influence of local ($\Gamma_{\rm S}$) and non-local ($\Gamma_{\rm C}$) tunneling processes on the quantum TUR for the CPS system is studied numerically for the current $J_{\rm S}$ flowing in lead S and presented in the top panel of Fig.~\ref{fig:CPS}. The violation of the classical and quantum TURs appears in three distinct regions around the points: $\Gamma_{\rm S}=\sqrt{5/3}\Gamma_{\rm N}$ and $\Gamma_{\rm C}=0$ (Reg.~1),  $\Gamma_{\rm S}=\Gamma_{\rm C}=\frac{1}{2}\sqrt{5/3}\Gamma_{\rm N}$ (Reg.~2), and $\Gamma_{\rm C}=\sqrt{5/3}\Gamma_{\rm N}$ and $\Gamma_{\rm S}=0$ (Reg.~3). 
For $\epsilon_{\rm L}=\epsilon_{\rm R}=0$ the current $J_{\rm S}$ and the noise $\mathcal{S}_{\rm S}$ are invariant under the exchange $\Gamma_{\rm S} \leftrightarrow \Gamma_{\rm C}$. This transformation interchanges the local (LAR) and crossed (CAR) Andreev reflection current contributions, leaving their sum unchanged.

In the bottom panels of  Fig.~\ref{fig:CPS}, the violation of the quantum TUR as a function of the energy levels $\epsilon_{\eta}$ is presented for the three different regions. In Reg.~1 (Reg.~3) transport is mediated exclusively by LAR (CAR) process, and the Fano factor of the superconducting current $J_S$ reaches its minimum in a finite range around the resonant condition $\epsilon_{\eta}
= 0$ ($\epsilon_{\rm L} = -\epsilon_{\rm R}$), where the corresponding Andreev reflection is maximal. In Reg.~2, the combined contribution of the LAR and CAR processes enhances the quantum-TUR violation by roughly a factor of $2.25$ relative to Regs.~1 and 3  at finite energy levels around   $\epsilon_{\eta}=-\epsilon_{\bar \eta}\approx \Gamma_{\rm N}/2\sqrt{3}$ with $\bar{\eta}\ne \eta$.  
\begin{figure}
\centering
\includegraphics[width=\linewidth]{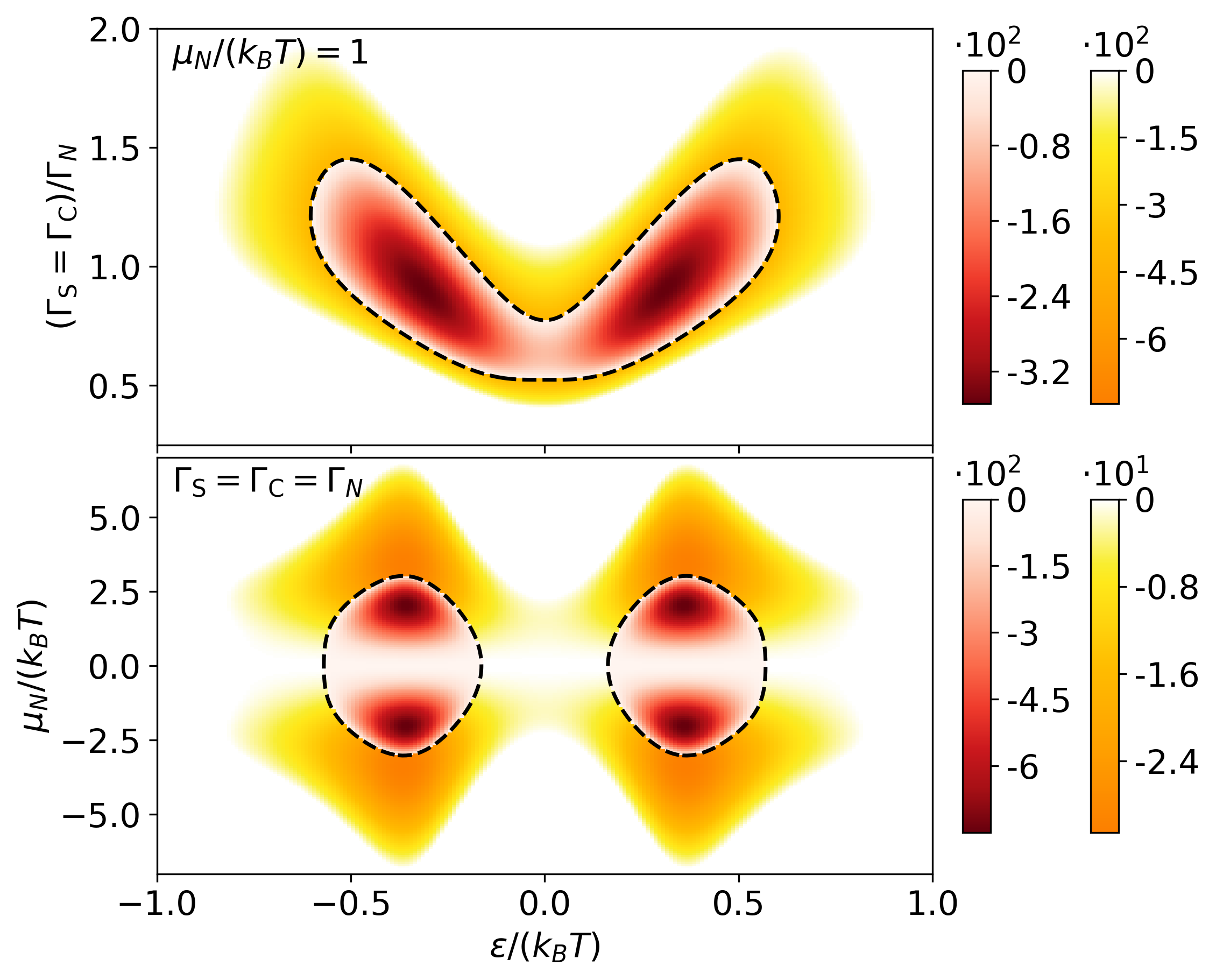}
\caption{Regions of violation of the classical (yellow) and quantum (red) TURs in the CPS system. The plotted quantities correspond to the left-hand side of Eq.~(1) where the corresponding inequality is violated. The dependence on the energy level $\varepsilon \equiv \varepsilon_L = -\varepsilon_R$ and the  tunneling rates $\Gamma_{\mathrm{S}}/\Gamma_{\mathrm{N}} = \Gamma_{\mathrm{C}}/\Gamma_{\mathrm{N}}$ is displayed in the top panel, while the dependence on the normalized chemical potential $\mu_{\mathrm{N}}/(k_{\mathrm{B}}T)$ is shown in the bottom panel. In the top panel, with $\mu_{\mathrm{N}} = k_{\mathrm{B}}T$ fixed, the largest departures from both bounds occur for nearly symmetric couplings, $\Gamma_{\mathrm{S}} = \Gamma_{\mathrm{C}} \approx \Gamma_{\mathrm{N}}$, at $|\varepsilon| \simeq \sqrt{5/12},\Gamma_{\mathrm{N}}$. In the bottom panel, fixing $\Gamma_{\mathrm{S}} = \Gamma_{\mathrm{C}} = \Gamma_{\mathrm{N}}$, the classical TUR reaches a minimum of $-0.3$ around $|\mu_{\mathrm{N}}|/(k_{\mathrm{B}}T) = 3.4$, while the quantum TUR attains $-0.075$ near $|\mu_{\mathrm{N}}|/(k_{\mathrm{B}}T) = 2$ (both at $|\varepsilon| \simeq \sqrt{5/12},\Gamma_{\mathrm{N}}$). Parameters (when fixed): $\Gamma_{\mathrm{N}} = 0.6\,k_{\mathrm{B}}T$, $\mu_{\mathrm{N}} = k_{\mathrm{B}}T$, and $\Gamma_{\mathrm{S}} = \Gamma_{\mathrm{C}} = \Gamma_{\mathrm{N}}$.}
\label{fig5}
\end{figure}

Figure~\ref{fig5} quantifies how the tunneling rates and the normal-lead chemical potential influence the degree of violation of the TURs along the CAR resonance line, $\varepsilon_L=-\varepsilon_R\equiv\varepsilon$. In the bottom panel, the shift between the classical and quantum TURs minima originates from the distinct chemical potential dependence entering the two bounds through the Fano factor: in the classical case, the violation grows approximately linearly with chemical potential, whereas the quantum bound, which incorporates the $\sinh(\mu_{\rm N}/k_{\rm B}T)$ dependence, saturates earlier.

\textit{Hybrid quantum bound}— For a two–terminal normal–superconducting junction in the $|\Delta|\!\to\!\infty$ limit with the superconducting terminal grounded we use the scattering approach to derive the hybrid quantum bound $\mathcal{B}_{\mathrm{H,qu}}(x)=2\,\mathrm{cosech}(2/x)$ for $x=2k_{B}|J|/(e\sigma)$ (see Supplemental Material~\cite{SM} for the derivation). 
Consequently, the hybrid quantum TUR takes the form
\begin{align}
\frac{\mathcal{S}}{2e|J|}\,\sinh\!\left(\frac{e\sigma}{k_{B}|J|}\right)-1\geq 0,
\label{eq:Hybrid_quantum_bound}
\end{align}
which is never violated in the systems considered here and is saturated only in the limit of vanishing current. 
A comparison of the hybrid quantum TUR of Eq.~\eqref{eq:Hybrid_quantum_bound} with the normal quantum TUR shows that the former can be obtained from the normal quantum TUR with the replacement $e\!\to\!2e$, reflecting the $2e$ charge transferred in Andreev processes. Eq.~\eqref{eq:Hybrid_quantum_bound} also holds for finite superconducting gap $\Delta$, although it is not saturated in this case.

\textit{Conclusions and Outlook---}In this work, we demonstrate that the macroscopic coherence arising from superconducting correlations in hybrid systems causes a violation of the recently formulated quantum TUR~\cite{Brandner2025}.
The classical TUR is also violated, and to a greater extent, even in the weak coupling regime, in contrast to what happens in normal (non-superconducting) systems.
We do this by calculating the Fano factor, using a Green's functions method, for the minimal system consisting of a QD coupled to a normal and a superconducting lead.
We find that the quantum TUR is violated in a smaller region of parameter space compared to the classical TUR, and it is more sensitive to decoherence. 
To understand the role of superconducting correlations, we introduce a decoherence probe. We find that this probe reduces the pairing amplitude in the QD, and for sufficiently strong decoherence, the TUR is no longer violated.

Macroscopic quantum coherence can also introduce non-local correlations. We find that in a Cooper pair splitter, in which two QDs are coupled to the superconducting lead, these correlations can lead to a stronger violation of the quantum TUR.

To account for macroscopic superconducting coherence, we have derived a hybrid quantum TUR that captures subgap Andreev transport in two-terminal normal–superconducting junctions in the $|\Delta|\!\to\!\infty$ limit. In the setups studied here, the bound is saturated in the limit of vanishing current.

\textit{Acknowledgments---}We acknowledge support from CONICET and Universidad de Buenos aires, Grant No. UBACyT (20020130100406BA) and ANPCyT (PICT-2019-04349 and PICT-2021-01288) (F.M.), the European Union under the Horizon Europe research and innovation programme (Marie Skłodowska-Curie grant agreement no. 101148213, EATTS) (N.S.), MUR-PRIN 2022 - Grant No. 2022B9P8LN - (PE3)-Project NEThEQS “Non-equilibrium coherent
thermal effects in quantum systems” in PNRR Mission 4 - Component 2 - Investment 1.1 “Fondo per il Programma Nazionale di Ricerca e Progetti di Rilevante Interesse
Nazionale (PRIN)” (F.T.), PNRR MUR project
PE0000023-NQSTI (R.F), the Royal Society through the International Exchanges between the UK and Italy (Grants No.
IEC R2 192166) (F.T), and the European Union (ERC, RAVE, 101053159) (R.F). M.G aknwoledges the generous hospitality of the International Centre of Theoretical Physics, Trieste, Italy.

%

	\clearpage
	\newpage
	
	\setcounter{section}{0}
	\setcounter{equation}{0}
	\setcounter{figure}{0}
	\setcounter{table}{0}
	
	\renewcommand{\thetable}{S\arabic{table}}
	\renewcommand{\theequation}{S\arabic{equation}}
	\renewcommand{\thefigure}{S\arabic{figure}}
	\renewcommand{\bibnumfmt}[1]{[S#1]}
	\renewcommand{\citenumfont}[1]{S#1}
	
	\onecolumngrid
	
\setcounter{section}{0}
\setcounter{equation}{0}
\setcounter{figure}{0}
\setcounter{table}{0}

\setcounter{page}{1}

\renewcommand{\thetable}{S\arabic{table}}
\renewcommand{\theequation}{S\arabic{equation}}
\renewcommand{\thefigure}{S\arabic{figure}}
\renewcommand{\bibnumfmt}[1]{[S#1]}
\renewcommand{\citenumfont}[1]{S#1}

\onecolumngrid

\begin{center}
	\textbf{\Large Supplemental Material for: Quantum Thermodynamic Uncertainty Relation and Macroscopic Superconducting Coherence}
	
	\bigskip
	
	Franco Mayo,$^{1,2,3}$ Nahual Sobrino,$^{3}$ Rosario Fazio,$^{3,4}$ Fabio Taddei,$^{5}$ and Michele Governale $^{6}$
	
	$^1$\!{\it Universidad de Buenos Aires, Facultad de Ciencias Exactas y Naturales, Departamento de F\'\i sica, Buenos Aires, Argentina}
	
	$^2$\!{\it CONICET-Universidad de Buenos Aires, Instituto de F\'\i sica de Buenos Aires, Buenos Aires, Argentina}
	
	$^3$\!{\it The Abdus Salam International Center for Theoretical Physics, Strada Costiera 11, 34151 Trieste, Italy}
	
	$^4$\!{\it Dipartimento di Fisica, Universit\`a di Napoli ``Federico II”, Monte S. Angelo, I-80126 Napoli, Italy}
	
	$^5$\!{\it NEST, Istituto Nanoscienze-CNR and Scuola Normale Superiore, Piazza San Silvestro 12, I-56127 Pisa, Italy}
	
	$^6$\!{\it School of Chemical and Physical Sciences and MacDiarmid Institute for Advanced Materials and Nanotechnology,
		Victoria University of Wellington, PO Box 600, Wellington 6140, New Zealand}
	
	\bigskip

\end{center}

\onecolumngrid

\appendix
\section*{Current and Noise for Hybrid Superconducting systems}

In this Supplemental Material we present all the necessary technical details on the derivations of the quantities needed for calculating the TUR in a quantum dot coupled to a superconducting lead and two normal leads, and in the Cooper pair splitter coupled to a superconducting lead and two normal leads. 

\subsection*{Single Quantum Dot}

For the system with a single quantum dot, we need to calculate the current leaving the normal lead as well as its noise. The current flowing from a normal lead $\mathrm{N}$ into the dot is defined as $J_{\rm N} = -e\langle \frac{d N_{\rm N}}{dt}\rangle = i \frac{e}{\hbar}\left[N_{\rm N}, H_{\rm tunn, N}\right]$, where
\begin{equation*}
	H_{\rm tunn, N} = V_{\rm N} \sum_{k, \sigma} \left(c^\dagger_{\rm{N} k\sigma}d_\sigma + d^\dagger_\sigma c_{\rm{N} k\sigma}\right),
\end{equation*}
and $N = \sum_k \boldsymbol{\Psi^\dagger_{\rm{N} k} \tau_3 \Psi_{\rm{N} k}}$ is written in Nambu representation, where $\boldsymbol{\Psi_{{\rm N} k}} = (c_{{\rm N} k\uparrow}, c^\dagger_{{\rm N} -k \downarrow})^\intercal$ and $\boldsymbol{\Phi} = (d_\uparrow, d^\dagger_\downarrow)^\intercal$. The expression for the current leaving the normal lead is therefore:
\begin{equation}
	J_{\rm N}(t) = -\frac{e}{\hbar}\sum_k i\,\boldsymbol{\Psi^\dagger_{{\rm N} k}(t)\tau_3\mathbf{V}_{\rm N}\Phi(t)} + \mathrm{H.c.},
\end{equation}
and $\mathbf{V}_{\rm N} = \mathrm{diag}(V_{\rm N}, -V^*_{\rm N})$ is a matrix that contains the coupling strength between the normal lead and the QD.

Defining $(\boldsymbol{G^<_{D,\rm{N}k}(\omega)})_{nm}$ as the Fourier transform of $i\langle \boldsymbol{\Psi^\dagger_{\mathrm{N}k m}(0)\Phi_n(t)}\rangle$ we can rewrite the current as
\begin{equation*}
	J_{\rm N} = -2\,\frac{e}{\hbar}\,\mathrm{Re}\left\{\sum_k\int \frac{d\omega}{2\pi}\mathrm{Tr}\left[\tau_3\,\boldsymbol{V_{\rm N}}\boldsymbol{G^<_{D,{\rm N} k}(\omega)}\right]\right\}.
\end{equation*}
Using a Dyson equation we can rewrite $\boldsymbol{G^<_{D,\rm{N}k}(\omega)})$ as
\begin{equation*}
	\boldsymbol{G^<_{D,{\rm N} k}(\omega)} = \boldsymbol{G^R_D(\omega)V^\dagger_\eta g^<_{\mathrm{N} k}(\omega)} + \boldsymbol{G^<_D(\omega)V^\dagger_\eta g^A_{\mathrm{N} k}(\omega)},
\end{equation*}
where $\boldsymbol{G_D(\omega)}$ refers to the full dot's Green's function (GF) and $\boldsymbol{g_{\mathrm{N} k}(\omega)}$ to the free lead GF. 
Therefore, the current leaving the normal lead can be expressed in terms of the full GF of the QD 
as~\cite{wingreen1992_SM, pala2007nonequilibrium_SM}:
\begin{equation}
	J_{\rm N} = -\frac{e}{h}\int 
	d\omega
	\Gamma_{\rm N}\mathrm{Im}\left\{\mathrm{Tr}\left[\tau_3\left(2\boldsymbol{F^+_{\rm N}}(\omega)\boldsymbol{G^R_D(\omega)} + \boldsymbol{G^<_D(\omega)}\right)\right]\right\},
	\label{eq:j_N}
\end{equation}
where $\boldsymbol{G^R_D(\omega)}$ and $\boldsymbol{G^<_D(\omega)}$ are the retarded and lesser full GFs of the QD in Nambu representation~\cite{pala2007nonequilibrium_SM}.
The matrices $\boldsymbol{F^{\pm}_{{\rm N}}(\omega)}$ are defined in terms of the Fermi function of the leads as
\begin{equation*}
	\boldsymbol{F^{\pm}_{{\rm N}}(\omega)} = \begin{bmatrix} f^\pm(\omega - \mu_{\rm N}) & 0 \\
		0 & f^\pm(\omega+\mu_{\rm N})\end{bmatrix},
\end{equation*} 

where $f^+(\omega)\equiv f(\omega)$ and $f^-(\omega) \equiv 1-f(\omega)$ with  $f(\omega)$ being the Fermi function at temperature $T$.

The other ingredient needed for the TUR is the zero-frequency noise in lead N, which is defined as~\cite{footnote1_SM}
\begin{equation}
	\mathcal{S}_{\rm{NN}} = \frac{1}{2}\,\int dt\, \langle J_{\rm N} (t)J_{\rm N}(0) + J_{\rm N}(0)J_{\rm N}(t)\rangle - \langle J_{\rm N}\rangle \langle J_{\rm N}\rangle,
\end{equation}
To obtain the zero frequency noise, we start with the general formula at frequency $\omega$, as it makes the calculation simpler:
\begin{equation}
	\mathcal{S}_{\rm NN}(\omega) = \frac{1}{2}\,\int dt\, e^{i\omega t} \langle J_{\rm N}(t)J_{\rm N} + J_{\rm N}J_{\rm N}(t)\rangle - \langle J_{\rm N}\rangle \langle J_{\rm N}\rangle.
\end{equation}
Using expression \eqref{eq:j_N} for the current it is easy to show that
\begin{equation}
	\mathcal{S}_{\rm NN}(\omega) = \,\int dt\, e^{i\omega t} \mathrm{Re}\left\{\langle \Tord{J_{\rm N}(t)J_{\rm N}}\rangle\right\} - \langle J_{\rm N}\rangle \langle J_{\rm N}\rangle.
\end{equation}
Applying Wick's theorem and rearranging, we arrive at:
\begin{align*}
	\mathrm{Re}\left\{\langle J_{\rm N}(t)J_{{\rm N}}\rangle\right\} &= \frac{e^2}{\hbar}\,V_{\rm N}\,V_{\rm N}\mathrm{Re} \sum_{k,k'}\sum_{i,j}\Big\{-\langle {\Psi^\dagger_{{\rm N} k i}(t)\Phi_i(t)}\rangle \langle \Tord{\Psi^\dagger_{{\rm N} k' j}\Phi_j}\rangle - \langle \Tord{\Psi^\dagger_{{\rm N} k i}(t)\Phi_j}\rangle \langle \Tord{\Phi_i(t)\Psi^\dagger_{{\rm N} k' j}}\rangle \\ 
	&+ \langle \Psi^\dagger_{{\rm N} k i}(t)\Phi_i(t)\rangle \langle \Phi^\dagger_j \Psi_{{\rm N} k' j}\rangle + \langle \Tord{\Psi^\dagger_{{\rm N} k i}(t)\Psi_{{\rm N}k' j}}\rangle \langle \Tord{\Phi_i(t) \Phi^\dagger_j}\rangle\Big\} .
\end{align*}
The first and third term, the ones in which the brackets only contain operators at equal times, combined with the factors outside of the sums give rise to a term that is the product of the currents, so we get:
\begin{align*}
	\mathrm{Re}\left\{\langle J_{\rm N}(t)J_{{\rm N}}\rangle\right\} &= \langle J_{\rm N}(t)\rangle \langle J_{{\rm N}}\rangle + \frac{e^2}{\hbar}\,V_{\rm N}\,V_{{\rm N}}\mathrm{Re} \sum_{k,k'}\sum_{i,j}\{-\langle \Tord{\Psi^\dagger_{{\rm N} k i}(t)\Phi_j}\rangle \langle \Tord{\Psi^\dagger_{{\rm N} k' j}\Phi_i(t)}\rangle \\ 
	& + \langle \Tord{\Psi^\dagger_{{\rm N} k i}(t)\Psi_{{\rm N}k' j}}\rangle \langle \Tord{\Phi_i(t) \Phi^\dagger_j}\rangle\} .
\end{align*}
In order to get rid of the time ordering, we expand all terms as 
\begin{equation*}
	\Tord{A(t)B} = \Theta(t) A(t)B - \Theta(-t) BA(t),
\end{equation*}
and after regrouping we obtain:
\begin{align*}
	\mathrm{Re}\left\{\langle J_{\rm N}(t)J_{{\rm N}}\rangle\right\} &= \langle J_{\rm N}(t)\rangle \langle J_{{\rm N}}\rangle + \frac{e^2}{\hbar}\,V_{\rm N}\,V_{{\rm N}}\mathrm{Re} \sum_{k,k'}\sum_{i,j}\Big\{ -\langle {\Psi^\dagger_{{\rm N} k i}(t)\Phi_j}\rangle \langle {\Phi_i(t)}\Psi^\dagger_{{\rm N} k' j}\rangle \\ 
	&- \langle {\Psi_{{\rm N} k i}(t)\Phi^\dagger_j}\rangle \langle {\Phi^\dagger_i(t)\Psi_{{\rm N} k' j}}\rangle
	+ \langle {\Psi_{{\rm N} k i}(t)\Psi^\dagger_{{\rm N}k' j}}\rangle \langle {\Phi^\dagger_i(t) \Phi_j}\rangle\\
	&+ \langle {\Psi^\dagger_{{\rm N}k' j}\Psi_{{\rm N} k i}(t)}\rangle \langle {\Phi_j\Phi^\dagger_i(t)}\rangle\Big\} .
\end{align*}

The remaining terms can be expressed as products of GFs, as:
\begin{align*}
	\mathrm{Re}\left\{\langle J_{\rm N}(t)J_{{\rm N}}\rangle\right\} &= \langle J_{\rm N}(t)\rangle \langle J_{{\rm N}}\rangle + \,\frac{e^2}{\hbar}\mathrm{Re} \sum_{k,k'}\Big\{ -2\Tr{\boldsymbol{\tau_3\,\mathbf{V}_{\rm N}\,G^<_{D,{\rm N} k}(-t)\,\tau_3\,\mathbf{V}_{{\rm N}}\,G^>_{D,{\rm N} k'}(t)}} \\ 
	& + \Tr{\boldsymbol{\tau_3\,\mathbf{V}_{\rm N}\,G^>_{{\rm N} k',{\rm N} }(-t)\,\tau_3\,\mathbf{V}_{{\rm N}}\,G^<_{D}(t)}}
	+ \Tr{\boldsymbol{\tau_3\,\mathbf{V}_{\rm N}\,G^<_{{\rm N} k,{\rm N}k'}(-t)\,\tau_3\,\mathbf{V}_{{\rm N}}\,G^>_{D}(t)}}\Big\},
\end{align*}
where we have used that $(\boldsymbol{G^<_{D, {\rm N} k}}(t))_{ij} = i\langle \Psi^\dagger_j\Phi_i(t)\rangle$, $(\boldsymbol{G^<_{D}}(t))_{ij} = -i\langle \Phi^\dagger_i\Phi_j(t)\rangle$ and $(\boldsymbol{G^<_{{\rm N} k, {\rm N} k'}}(t))_{ij} = i\langle \Psi^\dagger_j\Psi_i(t)\rangle$. 
And therefore the noise spectral density is:
\begin{align}
	\mathcal{S}_{{\rm N} {\rm N}}(\omega) &=\frac{e^2}{\hbar} \int dt\,e^{i\omega t} \,e^2\mathrm{Re} \sum_{k,k'}\{ -2\Tr{\boldsymbol{\tau_3\,\mathbf{V}_{\rm N}\,G^<_{D,{\rm N} k}(-t)\,\tau_3\,\mathbf{V}_{{\rm N}}\,G^>_{D,{\rm N} k'}(t)}} \\ \nonumber
	& + \Tr{\boldsymbol{\tau_3\,\mathbf{V}_{\rm N}\,G^>_{{\rm N} k',{\rm N} k}(-t)\,\tau_3\,\mathbf{V}_{{\rm N}}\,G^<_{D}(t)}}
	+ \Tr{\boldsymbol{\tau_3\,\mathbf{V}_{\rm N}\,G^<_{{\rm N} k,{\rm N}k'}(-t)\,\tau_3\,\mathbf{V}_{{\rm N}}\,G^>_{D}(t)}}\}.
\end{align}
Rewriting the GFs in terms of their Fourier transforms we obtain:
\begin{align}
	\mathcal{S}_{{\rm N} {\rm N}}(\omega) &=\frac{e^2}{h} \int 
	d\omega'
	\mathrm{Re} \Big\{\sum_{k,k'} -2\Tr{\boldsymbol{\tau_3\,\mathbf{V}_{\rm N}\,G^<_{D,{\rm N} k}(\omega')\,\tau_3\,\mathbf{V}_{{\rm N}}\,G^>_{D,{\rm N} k'}(\omega'+\omega)}} \\ \nonumber
	& + \Tr{\boldsymbol{\tau_3\,\mathbf{V}_{\rm N}\,G^>_{{\rm N} k,{\rm N} k'}(\omega')\,\tau_3\,\mathbf{V}_{{\rm N}}\,G^<_{D}(\omega'+\omega)}}
	+ \Tr{\boldsymbol{\tau_3\,\mathbf{V}_{\rm N}\,G^<_{{\rm N} k,{\rm N}k'}(\omega')\,\tau_3\,\mathbf{V}_{{\rm N}}\,G^>_{D}(\omega'+\omega)}}\Big\}.
\end{align}
We now use Dyson equations to separate the GFs into full dot GF and free normal lead GF. And then we perform the summation over $k$ and $k'$ by assuming a constant normal density of states and replacing the sum with an integral, obtaining the following formula for the zero frequency noise:

\begin{align}
	\nonumber
	\mathcal{S}_{N N}(0) = &\frac{e^2}{h}\Bigg[\Gamma_N\int d\omega\mathrm{Re}\left\{\Tr{i\boldsymbol{F^+(\omega)G_D^>(\omega)} - i\boldsymbol{F^-(\omega)G_D^<(\omega)}}\right\} \\ \nonumber
	&\int d\omega\frac{\Gamma_N^2}{4}\mathrm{Re}\left\{\Tr{\boldsymbol{\tau_3} \left[2\boldsymbol{G_D^R(\omega)F^+(\omega)} - 2\boldsymbol{F^+(\omega)G_D^A(\omega)} + \boldsymbol{G^<_D(\omega)}\right]\boldsymbol{\tau_3}\boldsymbol{G_D^>(\omega)}}\right\}\\ \nonumber
	&- \int d\omega\frac{\Gamma_N^2}{4}\mathrm{Re}\left\{\Tr{\boldsymbol{\tau_3} \left[2\boldsymbol{G_D^R(\omega)F^-(\omega)} - 2\boldsymbol{F^-(\omega)G_D^A(\omega)} - \boldsymbol{G^>_D(\omega)}\right]\boldsymbol{\tau_3}\boldsymbol{G_D^<(\omega)}}\right\}\\ 
	&-2 \int d\omega \frac{\Gamma_N^2}{4}\mathrm{Re}\left\{\Tr{\boldsymbol{\tau_3} \left[2\boldsymbol{G_D^R(\omega)F^+(\omega)}  + \boldsymbol{G^<_D(\omega)}\right]\boldsymbol{\tau_3}\left[2\boldsymbol{G_D^R(\omega)F^-(\omega)}  - \boldsymbol{G^>_D(\omega)}\right]}\right\}\Bigg] .     
	\label{eq:S_NN_formula}
\end{align}

In order to obtain the current and noise we need the GF of the QD. Here we provide the expressions for the case when the system is also coupled to a probe, i.e. a second normal lead with tunneling rate $\Gamma_{\rm P}$ (the S-QD-N case is obtained by setting $\Gamma_{\rm P} = 0$). The retarded and advanced GF can be written as:
\begin{equation}
	\boldsymbol{G^{R/A}_D(\omega)} = \frac{1}{(\omega \pm i\frac{\tilde{\Gamma}}{2})^2 - \epsilon_A^2} \begin{bmatrix}
		\omega + \epsilon \pm i\frac{\tilde{\Gamma}}{2} & -\frac{\Gamma_{\rm S}}{2} \\
		-\frac{\Gamma_{\rm S}}{2}  & \omega - \epsilon \pm i\frac{\tilde{\Gamma}}{2}
	\end{bmatrix}\,,
	\nonumber
	\label{eq:g_ra}
\end{equation}
where $\tilde{\Gamma} = \Gamma_{\rm N} + \Gamma_{\rm P}$ is the sum of tunneling rates for the normal lead and probe and $\epsilon_A=\sqrt{\epsilon^2+\frac{\Gamma_{\rm S}^2}{4}}$. The lesser and greater GF can be obtained by means of the Dyson equations~\cite{haug2008quantum_SM}
\begin{equation*}
	\boldsymbol{G^{\lessgtr}} = (1 + \boldsymbol{\Sigma^R\,G^R})\,\boldsymbol{G^{\lessgtr}_0}\,(1 + \boldsymbol{G^A\,\Sigma^A}) + \boldsymbol{G^R\Sigma^{\lessgtr}G^A},
\end{equation*}
where the self-energies are: $\boldsymbol{\Sigma^<(\omega)} = i\Gamma_{\rm N}\,\boldsymbol{F_{\rm N}^+(\omega)} + i\Gamma_{\rm P}\,\boldsymbol{F_{\rm P}^+(\omega)}$, $\boldsymbol{\Sigma^>(\omega)}=-i\Gamma_{\rm N}\,\boldsymbol{F_{\rm N}^-(\omega)} - i\Gamma_{\rm P}\,\boldsymbol{F_{\rm P}^-(\omega)}$, $\boldsymbol{\Sigma^R} = -i\tilde{\Gamma}/2$ and $\boldsymbol{\Sigma^A}=i\tilde{\Gamma}/2$.
Replacing this GFs in Eqs.~\eqref{eq:j_N} and \ref{eq:S_NN_formula} we can obtain the following expressions for the current leaving the normal lead and its noise respectively:
\begin{align}
	&J\equiv J_{\rm N} = - \frac{2e}{h}\,\int\,d\omega
	\,R_A(\omega )\,
	\Tr \left[\boldsymbol{\tau_3 F^+_{\rm N}(\omega)} \right]\, 
	\label{eq:JSDOTN}\\
	&\mathcal{S}=\mathcal{S}_{\rm NN} = \frac{4e^2}{h}\int d\omega \,\bigg[ R_A(\omega) \Tr \left[\boldsymbol{F^+_{\rm N}(\omega) F^-_{\rm N}(\omega)} \right]
	+  R_A(\omega)(1-R_A(\omega))\left(\Tr \left[\boldsymbol{\tau_3 F^+_{\rm N}(\omega)} \right]\right)^2\bigg]
	,
	\label{eq:SSDOTN}
\end{align}
with $R_A(\omega)$ being the Andreev reflection probability~\cite{splettstoesser2007pumping_SM}:
\begin{equation}
	R_A(\omega) = \pi\, \frac{\Gamma_{\rm N} \,\frac{\Gamma_{\rm S}^2}{4}}{\omega^2 + \epsilon_A^2 + \frac{\Gamma_{\rm N}^2}{4}}
	\left[L_{\Gamma_{\rm N}}(\omega-\epsilon_A) + L_{\Gamma_{\rm N}}(\omega+\epsilon_A)\right],
	\label{eq:AndrevReflection}
	\nonumber
\end{equation}
where $L_\Gamma(x) = \frac{1}{\pi} \frac{\Gamma/2}{x^2 + \Gamma^2/4}$ and $\epsilon_A=\sqrt{\epsilon^2+\frac{\Gamma_{\rm S}^2}{4}}$.

Since the normal current $J_{\rm N}$ is the only dissipative current in the system, the entropy-production rate is simply given by $\sigma=-\mu_{\rm N} J_{\rm N}/(eT)$, where $-\mu_{\rm N}/e$ is the voltage applied to the normal lead.

\subsection*{Cooper Pair Splitter}
For the Cooper pair splitter system, the retarded and advanced GFs can be written as: 
\begin{equation}
	\boldsymbol{G^{R/A}_{\rm CPS}(\omega)} =  \begin{bmatrix}
		\omega - \epsilon_L \pm i\frac{\Gamma_{\rm L}}{2} & \frac{\Gamma_{\rm S}}{2} & 0 &\frac{\Gamma_{\rm C}}{2} \\
		\frac{\Gamma_{\rm S}}{2}  & \omega + \epsilon_L \pm i\frac{\Gamma_{\rm L}}{2} & \frac{\Gamma_{\rm C}}{2} & 0 \\
		0 & \frac{\Gamma_{\rm C}}{2} &	\omega - \epsilon_R \pm i\frac{\Gamma_{\rm R}}{2}  &\frac{\Gamma_{\rm S}}{2} \\
		\frac{\Gamma_{\rm C}}{2} & 0 & \frac{\Gamma_{\rm S}}{2}& \omega + \epsilon_R \pm i\frac{\Gamma_{\rm R}}{2} 
	\end{bmatrix}^{-1}\,.
	\label{eq:g_ra_CPS}
\end{equation}
In the symmetrically coupled situation ($\Gamma_{\rm L}=\Gamma_{\rm R}\equiv\Gamma_{\rm N}$), expressions \eqref{eq:j_N} and \eqref{eq:S_NN_formula} remain valid for describing the current and noise in the normal leads of the CPS, provided that all Green’s function contributions are taken from the effective submatrix $\boldsymbol G_{\eta\eta}$ associated with lead $\eta$ in Eq.~\eqref{eq:g_ra_CPS}. It is instructive to express the total current in a normal lead $\eta$ using the scattering-matrix formalism as a sum of the independent Andreev reflection and cross Andreev reflection transport processes \cite{cao2015currents_SM}
\begin{align*}
	J_\eta= -\frac{2e}{h}\int d\omega 
	\left[ \left(f^{+}(\omega-\mu_\eta)- f^{-}(-\omega-\mu_\eta)\right)\bigl|s_{\eta\eta}^{eh}(\omega)\bigr|^{2}+ \left(f^{+}(\omega-\mu_\eta)- f^{-}(-\omega-\mu_{\bar\eta})\right)|s_{\eta\bar\eta}^{eh}(\omega)\bigr|^{2}\right]\,,
\end{align*}
where
$s_{\alpha\beta}^{mn}(\omega)=i\delta_{\alpha\beta}\delta_{mn}+\sqrt{\Gamma_{\alpha}\Gamma_{\beta}}\boldsymbol G^r_{\alpha\beta;mn}$ are the scattering amplitudes for particle type \(n\) in lead \(\beta\) to scatter into lead \(\alpha\) as type \(m\).
In the superconducting lead of the CPS system, the total noise is given by $\mathcal{S}_{\rm S} = \mathcal{S}_{\rm LL} + \mathcal{S}_{\rm RR} + 2\mathcal{S}_{\rm LR}$. The cross-term $\mathcal{S}_{\rm LR}$ can be expressed as \cite{anantram1996current_SM,cao2015currents_SM}
\begin{align}
	\mathcal{S}_{\eta\nu}(0)
	&=\frac{e^{2}}{h}\int\! d\omega\;     
	\mathrm{Re}\Bigl\{\mathrm{Tr}\Bigl\{
	\delta_{\eta\nu}\Bigl(
	\boldsymbol{G^{<}_{\eta\eta}(\omega)}\,\boldsymbol{\tau_{3}}\,\boldsymbol{\Sigma^{>}_{\eta}(\omega)}\,\boldsymbol{\tau_{3}}
	+ \boldsymbol{\tau_{3}}\,\boldsymbol{\Sigma^{<}_{\eta}(\omega)}\,\boldsymbol{\tau_{3}}\,\boldsymbol{G^{>}_{\eta\eta}(\omega)}
	\Bigr)\nonumber\\[4pt]
	&
	+ \boldsymbol{G^{<}_{\eta\nu}(\omega)}\,\boldsymbol{\tau_{3}}\,
	\Bigl(
	\boldsymbol{\Sigma^{r}_{\nu}(\omega)}\,\boldsymbol{G^{r}_{\nu\eta}(\omega)}\,\boldsymbol{\Sigma^{>}_{\eta}(\omega)}
	+ \boldsymbol{\Sigma^{r}_{\nu}(\omega)}\,\boldsymbol{G^{>}_{\nu\eta}(\omega)}\,\boldsymbol{\Sigma^{a}_{\eta}(\omega)}
	+ \boldsymbol{\Sigma^{>}_{\nu}(\omega)}\,\boldsymbol{G^{a}_{\nu\eta}(\omega)}\,\boldsymbol{\Sigma^{a}_{\eta}(\omega)}
	\Bigr)\,\boldsymbol{\tau_{3}}\nonumber\\[4pt]
	&
	- \boldsymbol{\tau_{3}}\,
	\Bigl(
	\boldsymbol{\Sigma^{r}_{\eta}(\omega)}\,\boldsymbol{G^{<}_{\eta\nu}(\omega)}
	+ \boldsymbol{\Sigma^{<}_{\eta}(\omega)}\,\boldsymbol{G^{a}_{\eta\nu}(\omega)}
	\Bigr)\,\boldsymbol{\tau_{3}}\,
	\Bigl(
	\boldsymbol{\Sigma^{r}_{\nu}(\omega)}\,\boldsymbol{G^{>}_{\nu\eta}(\omega)}
	+ \boldsymbol{\Sigma^{>}_{\nu}(\omega)}\,\boldsymbol{G^{a}_{\nu\eta}(\omega)}
	\Bigr)\nonumber\\[4pt]
	&
	- \Bigl(
	\boldsymbol{G^{r}_{\eta\nu}(\omega)}\,\boldsymbol{\Sigma^{<}_{\nu}(\omega)}
	+ \boldsymbol{G^{<}_{\eta\nu}(\omega)}\,\boldsymbol{\Sigma^{a}_{\nu}(\omega)}
	\Bigr)\,
	\boldsymbol{\tau_{3}}\,
	\Bigl(
	\boldsymbol{G^{r}_{\nu\eta}(\omega)}\,\boldsymbol{\Sigma^{>}_{\eta}(\omega)}
	+ \boldsymbol{G^{>}_{\nu\eta}(\omega)}\,\boldsymbol{\Sigma^{a}_{\eta}(\omega)}
	\Bigr)\,\boldsymbol{\tau_{3}}\nonumber\\[4pt]
	&
	+ \boldsymbol{\tau_{3}}\,
	\Bigl(
	\boldsymbol{\Sigma^{r}_{\eta}(\omega)}\,\boldsymbol{G^{r}_{\eta\nu}(\omega)}\,\boldsymbol{\Sigma^{<}_{\nu}(\omega)}
	+ \boldsymbol{\Sigma^{r}_{\eta}(\omega)}\,\boldsymbol{G^{<}_{\eta\nu}(\omega)}\,\boldsymbol{\Sigma^{a}_{\nu}(\omega)}
	+ \boldsymbol{\Sigma^{<}_{\eta}(\omega)}\,\boldsymbol{G^{a}_{\eta\nu}(\omega)}\,\boldsymbol{\Sigma^{a}_{\nu}(\omega)}
	\Bigr)\,\boldsymbol{\tau_{3}}\,\boldsymbol{G^{>}_{\nu\eta}(\omega)}
	\Bigr\}\Bigr\},
\end{align}
with the GFs and self-energies as defined above, where $\,G^{x}_{\eta\nu}\;(x=R,A,<,>)$ is obtained by selecting the rows and columns
associated with leads $\eta$ and $\nu$ of the full $4\times4$ Green’s function and $\,\Sigma^{x}_{\eta}\;(x=R,A,<,>)$ the $2\times2$ matrix related to lead $\eta$. 
In this system, the entropy-production rate is given by $\sigma=-(\mu_{\rm L} J_{\rm L}+\mu_{\rm R} J_{\rm R})/(eT)$.

\section*{TUR for Hybrid Normal Superconducting System}
\begin{figure}
	\centering
	\includegraphics[width=0.5\linewidth]{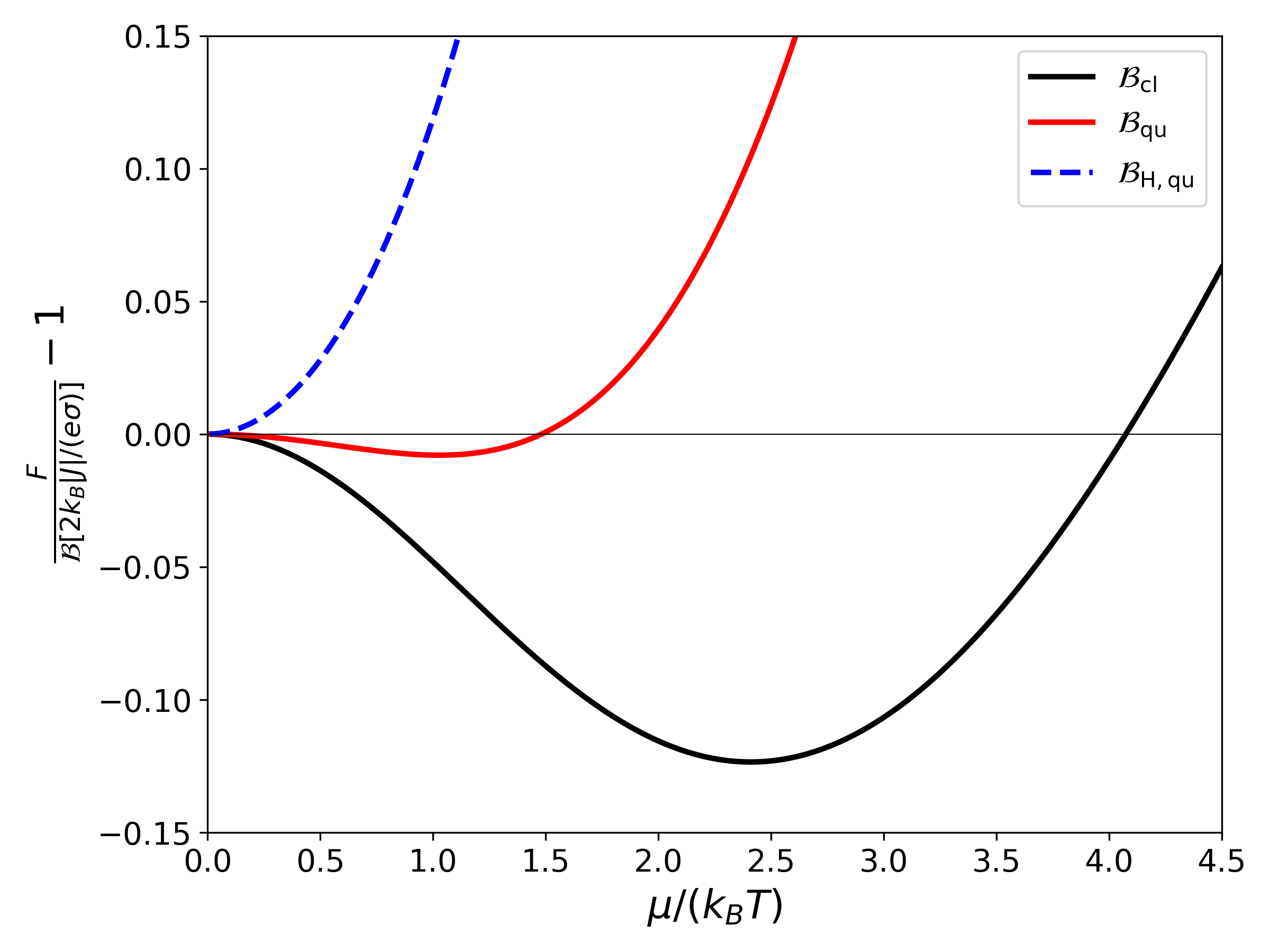}
	\caption{The Classical, Quantum and Hybrid Quantum bound, for a quantum dot coupled to one normal and one superconducting lead. The parameters are: $\Gamma_N = 0.1 k_B T$, $\Gamma_S = \sqrt{5/3} \Gamma_N$, $\epsilon = 0$. The Hybrid Quantum bound is only saturated for $\mu\to 0$.}
	\label{fig:TUR_comparison}
\end{figure}

We consider a two-terminal setup with one normal lead and one superconducting lead, and we take the limit of infinite superconducting gap. Furthermore, we allow for multiple propagating modes and assume the superconductor is grounded, that is $\mu_{\text{S}}=0$.  
For convenience, in this section we use the following notation for the Fermi functions 
$ f^{e/h}(\omega)=1/\left(\exp(\frac{\omega\mp\mu}{k_B T})+1\right)$, and $\mu$ and $T$ are, respectively, the chemical potential and the temperature of the normal lead. 
The current and the noise~\cite{beenakker1994_SM,anantram1996current_SM} can be written as 
\begin{align*}
	J=&-\frac{2e}{h}\int\, d\omega \sum_n \mathcal{R}_n (\omega)\left(f^{e}(\omega)-f^{h}(\omega)\right)\\    
	\mathcal{S}=&\frac{4 e^2}{h}\int\, d\omega \sum_n \Big\{\mathcal{R}_n (\omega)\left[f^{e}(\omega)(1-f^{e}(\omega))+f^{h}(\omega)(1-f^{h}(\omega))\right]+\mathcal{R}_n (\omega)(1-\mathcal{R}_n (\omega))
	\left(f^{e}(\omega)-f^{h}(\omega) \right)^2\Big\},
\end{align*}
where $\mathcal{R}_n(\omega)$ are the eigenvalues of the of the matrix $\boldsymbol{r}_{\rm eh}\boldsymbol{r}_{\rm eh}^\dagger$, $\boldsymbol{r}_{\rm eh}$ being the matrix containing the Andreev reflection amplitudes.
This is just the multimode generalization of Eqs.~(\ref{eq:JSDOTN}) and (\ref{eq:SSDOTN}). 
The noise can be written as $\mathcal{S}=\mathcal{S}_{\text{th}}+\mathcal{S}_{\text{sh}}$, where $\mathcal{S}_{\text{th}}$ is given 
by 
\begin{align}
	\mathcal{S}_{\text{th}}&=\frac{4 e^2}{h}\int\, d\omega \sum_n\, \mathcal{R}_n (\omega)\left[f^{e}(\omega)(1-f^{e}(\omega))+f^{h}(\omega)(1-f^{h}(\omega))\right]\\
	&=\frac{4 e^2}{h}\int\, d\omega \sum_n\,\mathcal{R}_n (\omega)\left[g^{ee}(\omega)+g^{hh}(\omega)\right],
\end{align}
where we have introduced the following notation
\begin{align*}
	g^{\alpha\beta}(\omega)=   f^{\alpha}(\omega)(1-f^{\beta}(\omega)),
\end{align*}
with $\alpha, \beta\in \{e,\ h\}$.
We start by noticing that 
\begin{align}
	\frac{V}{T}=-\frac{\mu}{e T}= -\frac{k_B}{2 e}\log\left(\frac{1-f^{h}(\omega)}{f^{h}(\omega)}\,\frac{f^{e}(\omega)}{1-f^{e}(\omega)}\right)=-\frac{k_B}{2 e}\log\left(\frac{g^{eh}(\omega)}{g^{he}(\omega)}\right).
\end{align}
The rate of entropy production is given by 
\begin{align}
	\sigma=\frac{J V}{T}=\frac{k_B}{h}\int\, dE \sum_n\, \mathcal{R}_n (\omega)\left(f^{e}(\omega)-f^{h}(\omega)\right)\log\left(\frac{g^{eh}(\omega)}{g^{he}(\omega)}\right).
\end{align}
We follow closely Brandner and Saito\cite{Brandner-2025_SM} and define the quantity $X^{eh}(\omega)$:
\begin{align}
	X^{eh}(\omega)=\frac{f^{e}(\omega)-f^{h}(\omega)}{2\sqrt{g^{ee}(\omega)g^{hh}(\omega))}}.
\end{align}
In terms of this quantity, the rate of entropy production reads 
\begin{align}
	\label{eq:sigm1}
	\sigma=\frac{4 k_B}{h}\int\, d\omega \sum_n\, \mathcal{R}_n (\omega) \sqrt{g^{ee}(\omega)g^{hh}(\omega)}\, \Phi(X^{eh}(\omega)), 
\end{align}
with $\Phi(x)=x\,\text{arcsinh}(x)$.
Due to the convexity of $\Phi(x)$, we have the following inequality
\begin{align}
	\Phi[X^{eh}(\omega)]\ge \Phi(X)+\Phi'(X)(X^{eh}(\omega)-X),
\end{align}
which holds for any $X$.
Substituting in Eq.~(\ref{eq:sigm1}), we obtain
\begin{align}
	\label{eq:sigma2}
	\sigma\ge\frac{4 k_B}{h}\int\, d\omega \sum_n\, \mathcal{R}_n (\omega) \sqrt{g^{ee}(\omega)g^{hh}(\omega)}\left[ \Phi(X)+\Phi'(X)(X^{eh}(\omega)-X)\right].
\end{align}
We now choose $X$ so that the term in $\Phi'$ vanishes, that is 
\begin{align*}
	X=\frac{e^2}{2N}\frac{1}{h}\int\, d\omega \sum_n\, \mathcal{R}_n (\omega)\left(f^{e}(\omega)-f^{h}(\omega)\right)=-e\frac{J}{4N}={e}\frac{|J|}{4N}
\end{align*}
with 
\begin{align*}
	N=\frac{e^2}{h}\int d\omega \sum_n\, \mathcal{R}_n (\omega)\sqrt{g^{ee}(\omega)g^{hh}(\omega)}.
\end{align*}
Substituting the value of $X$ given above into Eq.~(\ref{eq:sigma2}), yield
\begin{align}
	\sigma\ge \frac{k_B|J|}{e} \, \text{arcsinh}\left(\frac{e|J|}{4N}\right).
\end{align}
Finally, we can relate $N$ to the noise:
\begin{align*}
	4N=\frac{4 e^2}{h}\int d\omega \sum_n\, \mathcal{R}_n (\omega)\sqrt{g^{ee}(\omega)g^{hh}(\omega)}\le\frac{2 e^2}{h}\int d\omega \sum_n\, \mathcal{R}_n (\omega)\left[g^{ee}(\omega)+g^{hh}(\omega)\right]=\frac{\mathcal{S}^{th}}{2}\le\frac{\mathcal{S}}{2}. 
\end{align*}
The inequality for the entropy production can therefore be written as 
\begin{align*}
	\sigma\ge  \frac{ k_B|J|}{e}\,\text{arcsinh}\left(\frac{ 2 e  |J|}{\mathcal{S}}\right)
\end{align*}
or alternatively as  
\begin{align}
	\label{eq:newTUR}
	\frac{\mathcal{S}}{2 e |J|}\sinh\left(\frac{e \sigma}{k_B |J|}\right)\ge 1.
\end{align}
This is the key result of this section and a few comments are in order. 
It is interesting to compare the hybrid quantum TUR of Eq.~(\ref{eq:newTUR}) with the quantum TUR for the normal case 
\begin{align}
	\label{eq:quantumTUR}
	\frac{\mathcal{S}}{e |J|}\sinh\left(\frac{e \sigma}{2 k_B |J|}\right)\ge 1. \quad\quad\text{(normal quantum TUR)}
\end{align}
Formally, Eq.~(\ref{eq:newTUR}) can be obtained from Eq.~(\ref{eq:quantumTUR}) by the replacement $e\rightarrow 2e$. This enables a clear interpretation of the new bound. 
In Fig. \ref{fig:TUR_comparison} a comparison between  the Classical, Quantum and Hybrid Quantum bounds is shown for a quantum dot coupled to one normal and one superconducting lead. The Hybrid Quantum TUR is never saturated for the quantum-dot case except at zero current, that is, for $\mu\rightarrow 0$.

\begin{thebibliography}{46}%
\makeatletter
\providecommand \@ifxundefined [1]{%
	\@ifx{#1\undefined}
}%
\providecommand \@ifnum [1]{%
	\ifnum #1\expandafter \@firstoftwo
	\else \expandafter \@secondoftwo
	\fi
}%
\providecommand \@ifx [1]{%
	\ifx #1\expandafter \@firstoftwo
	\else \expandafter \@secondoftwo
	\fi
}%
\providecommand \natexlab [1]{#1}%
\providecommand \enquote  [1]{``#1''}%
\providecommand \bibnamefont  [1]{#1}%
\providecommand \bibfnamefont [1]{#1}%
\providecommand \citenamefont [1]{#1}%
\providecommand \href@noop [0]{\@secondoftwo}%
\providecommand \href [0]{\begingroup \@sanitize@url \@href}%
\providecommand \@href[1]{\@@startlink{#1}\@@href}%
\providecommand \@@href[1]{\endgroup#1\@@endlink}%
\providecommand \@sanitize@url [0]{\catcode `\\12\catcode `\$12\catcode `\&12\catcode `\#12\catcode `\^12\catcode `\_12\catcode `\%12\relax}%
\providecommand \@@startlink[1]{}%
\providecommand \@@endlink[0]{}%
\providecommand \url  [0]{\begingroup\@sanitize@url \@url }%
\providecommand \@url [1]{\endgroup\@href {#1}{\urlprefix }}%
\providecommand \urlprefix  [0]{URL }%
\providecommand \Eprint [0]{\href }%
\providecommand \doibase [0]{https://doi.org/}%
\providecommand \selectlanguage [0]{\@gobble}%
\providecommand \bibinfo  [0]{\@secondoftwo}%
\providecommand \bibfield  [0]{\@secondoftwo}%
\providecommand \translation [1]{[#1]}%
\providecommand \BibitemOpen [0]{}%
\providecommand \bibitemStop [0]{}%
\providecommand \bibitemNoStop [0]{.\EOS\space}%
\providecommand \EOS [0]{\spacefactor3000\relax}%
\providecommand \BibitemShut  [1]{\csname bibitem#1\endcsname}%
\let\auto@bib@innerbib\@empty
\bibitem [{\citenamefont {Barato}\ and\ \citenamefont {Seifert}(2015)}]{barato2015thermodynamic}%
\BibitemOpen
\bibfield  {author} {\bibinfo {author} {\bibfnamefont {A.~C.}\ \bibnamefont {Barato}}\ and\ \bibinfo {author} {\bibfnamefont {U.}~\bibnamefont {Seifert}},\ }\bibfield  {title} {\bibinfo {title} {Thermodynamic uncertainty relation for biomolecular processes},\ }\href@noop {} {\bibfield  {journal} {\bibinfo  {journal} {Phys. Rev. Lett.}\ }\textbf {\bibinfo {volume} {114}},\ \bibinfo {pages} {158101} (\bibinfo {year} {2015})}\BibitemShut {NoStop}%
\bibitem [{\citenamefont {Gingrich}\ \emph {et~al.}(2016)\citenamefont {Gingrich}, \citenamefont {Horowitz}, \citenamefont {Perunov},\ and\ \citenamefont {England}}]{gingrich2016dissipation}%
\BibitemOpen
\bibfield  {author} {\bibinfo {author} {\bibfnamefont {T.~R.}\ \bibnamefont {Gingrich}}, \bibinfo {author} {\bibfnamefont {J.~M.}\ \bibnamefont {Horowitz}}, \bibinfo {author} {\bibfnamefont {N.}~\bibnamefont {Perunov}},\ and\ \bibinfo {author} {\bibfnamefont {J.~L.}\ \bibnamefont {England}},\ }\bibfield  {title} {\bibinfo {title} {Dissipation bounds all steady-state current fluctuations},\ }\href@noop {} {\bibfield  {journal} {\bibinfo  {journal} {Phys. Rev. Lett.}\ }\textbf {\bibinfo {volume} {116}},\ \bibinfo {pages} {120601} (\bibinfo {year} {2016})}\BibitemShut {NoStop}%
\bibitem [{\citenamefont {Pietzonka}\ \emph {et~al.}(2016)\citenamefont {Pietzonka}, \citenamefont {Barato},\ and\ \citenamefont {Seifert}}]{pietzonka2016universal}%
\BibitemOpen
\bibfield  {author} {\bibinfo {author} {\bibfnamefont {P.}~\bibnamefont {Pietzonka}}, \bibinfo {author} {\bibfnamefont {A.~C.}\ \bibnamefont {Barato}},\ and\ \bibinfo {author} {\bibfnamefont {U.}~\bibnamefont {Seifert}},\ }\bibfield  {title} {\bibinfo {title} {Universal bounds on current fluctuations},\ }\href@noop {} {\bibfield  {journal} {\bibinfo  {journal} {Phys. Rev. E}\ }\textbf {\bibinfo {volume} {93}},\ \bibinfo {pages} {052145} (\bibinfo {year} {2016})}\BibitemShut {NoStop}%
\bibitem [{\citenamefont {Timpanaro}\ \emph {et~al.}(2019)\citenamefont {Timpanaro}, \citenamefont {Guarnieri}, \citenamefont {Goold},\ and\ \citenamefont {Landi}}]{timpanaro2019thermodynamic}%
\BibitemOpen
\bibfield  {author} {\bibinfo {author} {\bibfnamefont {A.~M.}\ \bibnamefont {Timpanaro}}, \bibinfo {author} {\bibfnamefont {G.}~\bibnamefont {Guarnieri}}, \bibinfo {author} {\bibfnamefont {J.}~\bibnamefont {Goold}},\ and\ \bibinfo {author} {\bibfnamefont {G.~T.}\ \bibnamefont {Landi}},\ }\bibfield  {title} {\bibinfo {title} {Thermodynamic uncertainty relations from exchange fluctuation theorems},\ }\href@noop {} {\bibfield  {journal} {\bibinfo  {journal} {Phys. Rev. Lett.}\ }\textbf {\bibinfo {volume} {123}},\ \bibinfo {pages} {090604} (\bibinfo {year} {2019})}\BibitemShut {NoStop}%
\bibitem [{\citenamefont {Seifert}(2019)}]{Seifert2019}%
\BibitemOpen
\bibfield  {author} {\bibinfo {author} {\bibfnamefont {U.}~\bibnamefont {Seifert}},\ }\bibfield  {title} {\bibinfo {title} {From stochastic thermodynamics to thermodynamic inference},\ }\href {https://doi.org/https://doi.org/10.1146/annurev-conmatphys-031218-013554} {\bibfield  {journal} {\bibinfo  {journal} {Annual Review of Condensed Matter Physics}\ }\textbf {\bibinfo {volume} {10}},\ \bibinfo {pages} {171} (\bibinfo {year} {2019})}\BibitemShut {NoStop}%
\bibitem [{\citenamefont {Horowitz}\ and\ \citenamefont {Gingrich}(2020)}]{horowitz2020thermodynamic}%
\BibitemOpen
\bibfield  {author} {\bibinfo {author} {\bibfnamefont {J.~M.}\ \bibnamefont {Horowitz}}\ and\ \bibinfo {author} {\bibfnamefont {T.~R.}\ \bibnamefont {Gingrich}},\ }\bibfield  {title} {\bibinfo {title} {Thermodynamic uncertainty relations constrain non-equilibrium fluctuations},\ }\href@noop {} {\bibfield  {journal} {\bibinfo  {journal} {Nature Physics}\ }\textbf {\bibinfo {volume} {16}},\ \bibinfo {pages} {15} (\bibinfo {year} {2020})}\BibitemShut {NoStop}%
\bibitem [{\citenamefont {Agarwalla}\ and\ \citenamefont {Segal}(2018)}]{agarwalla2018assessing}%
\BibitemOpen
\bibfield  {author} {\bibinfo {author} {\bibfnamefont {B.~K.}\ \bibnamefont {Agarwalla}}\ and\ \bibinfo {author} {\bibfnamefont {D.}~\bibnamefont {Segal}},\ }\bibfield  {title} {\bibinfo {title} {Assessing the validity of the thermodynamic uncertainty relation in quantum systems},\ }\href@noop {} {\bibfield  {journal} {\bibinfo  {journal} {Phys. Rev. B}\ }\textbf {\bibinfo {volume} {98}},\ \bibinfo {pages} {155438} (\bibinfo {year} {2018})}\BibitemShut {NoStop}%
\bibitem [{\citenamefont {Ehrlich}\ and\ \citenamefont {Schaller}(2021)}]{Ehrlich2021}%
\BibitemOpen
\bibfield  {author} {\bibinfo {author} {\bibfnamefont {T.}~\bibnamefont {Ehrlich}}\ and\ \bibinfo {author} {\bibfnamefont {G.}~\bibnamefont {Schaller}},\ }\bibfield  {title} {\bibinfo {title} {Broadband frequency filters with quantum dot chains},\ }\href {https://doi.org/10.1103/PhysRevB.104.045424} {\bibfield  {journal} {\bibinfo  {journal} {Phys. Rev. B}\ }\textbf {\bibinfo {volume} {104}},\ \bibinfo {pages} {045424} (\bibinfo {year} {2021})}\BibitemShut {NoStop}%
\bibitem [{\citenamefont {Gerry}\ and\ \citenamefont {Segal}(2022)}]{gerry2022absence}%
\BibitemOpen
\bibfield  {author} {\bibinfo {author} {\bibfnamefont {M.}~\bibnamefont {Gerry}}\ and\ \bibinfo {author} {\bibfnamefont {D.}~\bibnamefont {Segal}},\ }\bibfield  {title} {\bibinfo {title} {Absence and recovery of cost-precision tradeoff relations in quantum transport},\ }\href@noop {} {\bibfield  {journal} {\bibinfo  {journal} {Phys. Rev. B}\ }\textbf {\bibinfo {volume} {105}},\ \bibinfo {pages} {155401} (\bibinfo {year} {2022})}\BibitemShut {NoStop}%
\bibitem [{\citenamefont {Guarnieri}\ \emph {et~al.}(2019)\citenamefont {Guarnieri}, \citenamefont {Landi}, \citenamefont {Clark},\ and\ \citenamefont {Goold}}]{Guarnieri2019}%
\BibitemOpen
\bibfield  {author} {\bibinfo {author} {\bibfnamefont {G.}~\bibnamefont {Guarnieri}}, \bibinfo {author} {\bibfnamefont {G.~T.}\ \bibnamefont {Landi}}, \bibinfo {author} {\bibfnamefont {S.~R.}\ \bibnamefont {Clark}},\ and\ \bibinfo {author} {\bibfnamefont {J.}~\bibnamefont {Goold}},\ }\bibfield  {title} {\bibinfo {title} {Thermodynamics of precision in quantum nonequilibrium steady states},\ }\href {https://doi.org/10.1103/PhysRevResearch.1.033021} {\bibfield  {journal} {\bibinfo  {journal} {Phys. Rev. Res.}\ }\textbf {\bibinfo {volume} {1}},\ \bibinfo {pages} {033021} (\bibinfo {year} {2019})}\BibitemShut {NoStop}%
\bibitem [{\citenamefont {Kheradsoud}\ \emph {et~al.}(2019)\citenamefont {Kheradsoud}, \citenamefont {Dashti}, \citenamefont {Misiorny}, \citenamefont {Potts}, \citenamefont {Splettstoesser},\ and\ \citenamefont {Samuelsson}}]{Kheradsoud2019}%
\BibitemOpen
\bibfield  {author} {\bibinfo {author} {\bibfnamefont {S.}~\bibnamefont {Kheradsoud}}, \bibinfo {author} {\bibfnamefont {N.}~\bibnamefont {Dashti}}, \bibinfo {author} {\bibfnamefont {M.}~\bibnamefont {Misiorny}}, \bibinfo {author} {\bibfnamefont {P.~P.}\ \bibnamefont {Potts}}, \bibinfo {author} {\bibfnamefont {J.}~\bibnamefont {Splettstoesser}},\ and\ \bibinfo {author} {\bibfnamefont {P.}~\bibnamefont {Samuelsson}},\ }\bibfield  {title} {\bibinfo {title} {Power, efficiency and fluctuations in a quantum point contact as steady-state thermoelectric heat engine},\ }\bibfield  {journal} {\bibinfo  {journal} {Entropy}\ }\textbf {\bibinfo {volume} {21}},\ \href {https://doi.org/10.3390/e21080777} {10.3390/e21080777} (\bibinfo {year} {2019})\BibitemShut {NoStop}%
\bibitem [{\citenamefont {Proesmans}\ and\ \citenamefont {Horowitz}(2019)}]{Proesmans2019}%
\BibitemOpen
\bibfield  {author} {\bibinfo {author} {\bibfnamefont {K.}~\bibnamefont {Proesmans}}\ and\ \bibinfo {author} {\bibfnamefont {J.~M.}\ \bibnamefont {Horowitz}},\ }\bibfield  {title} {\bibinfo {title} {Hysteretic thermodynamic uncertainty relation for systems with broken time-reversal symmetry},\ }\href {https://doi.org/10.1088/1742-5468/ab14da} {\bibfield  {journal} {\bibinfo  {journal} {Journal of Statistical Mechanics: Theory and Experiment}\ }\textbf {\bibinfo {volume} {2019}},\ \bibinfo {pages} {054005} (\bibinfo {year} {2019})}\BibitemShut {NoStop}%
\bibitem [{\citenamefont {Liu}\ and\ \citenamefont {Segal}(2019)}]{liu2019thermodynamic}%
\BibitemOpen
\bibfield  {author} {\bibinfo {author} {\bibfnamefont {J.}~\bibnamefont {Liu}}\ and\ \bibinfo {author} {\bibfnamefont {D.}~\bibnamefont {Segal}},\ }\bibfield  {title} {\bibinfo {title} {Thermodynamic uncertainty relation in quantum thermoelectric junctions},\ }\href@noop {} {\bibfield  {journal} {\bibinfo  {journal} {Phys. Rev. E}\ }\textbf {\bibinfo {volume} {99}},\ \bibinfo {pages} {062141} (\bibinfo {year} {2019})}\BibitemShut {NoStop}%
\bibitem [{\citenamefont {Pietzonka}\ and\ \citenamefont {Seifert}(2018)}]{Pietzonka2018}%
\BibitemOpen
\bibfield  {author} {\bibinfo {author} {\bibfnamefont {P.}~\bibnamefont {Pietzonka}}\ and\ \bibinfo {author} {\bibfnamefont {U.}~\bibnamefont {Seifert}},\ }\bibfield  {title} {\bibinfo {title} {Universal trade-off between power, efficiency, and constancy in steady-state heat engines},\ }\href {https://doi.org/10.1103/PhysRevLett.120.190602} {\bibfield  {journal} {\bibinfo  {journal} {Phys. Rev. Lett.}\ }\textbf {\bibinfo {volume} {120}},\ \bibinfo {pages} {190602} (\bibinfo {year} {2018})}\BibitemShut {NoStop}%
\bibitem [{\citenamefont {Prech}\ \emph {et~al.}(2023)\citenamefont {Prech}, \citenamefont {Johansson}, \citenamefont {Nyholm}, \citenamefont {Landi}, \citenamefont {Verdozzi}, \citenamefont {Samuelsson},\ and\ \citenamefont {Potts}}]{Prech2023}%
\BibitemOpen
\bibfield  {author} {\bibinfo {author} {\bibfnamefont {K.}~\bibnamefont {Prech}}, \bibinfo {author} {\bibfnamefont {P.}~\bibnamefont {Johansson}}, \bibinfo {author} {\bibfnamefont {E.}~\bibnamefont {Nyholm}}, \bibinfo {author} {\bibfnamefont {G.~T.}\ \bibnamefont {Landi}}, \bibinfo {author} {\bibfnamefont {C.}~\bibnamefont {Verdozzi}}, \bibinfo {author} {\bibfnamefont {P.}~\bibnamefont {Samuelsson}},\ and\ \bibinfo {author} {\bibfnamefont {P.~P.}\ \bibnamefont {Potts}},\ }\bibfield  {title} {\bibinfo {title} {Entanglement and thermokinetic uncertainty relations in coherent mesoscopic transport},\ }\href {https://doi.org/10.1103/PhysRevResearch.5.023155} {\bibfield  {journal} {\bibinfo  {journal} {Phys. Rev. Res.}\ }\textbf {\bibinfo {volume} {5}},\ \bibinfo {pages} {023155} (\bibinfo {year} {2023})}\BibitemShut {NoStop}%
\bibitem [{\citenamefont {Ptaszy\ifmmode~\acute{n}\else \'{n}\fi{}ski}(2018)}]{Ptaszynski2018}%
\BibitemOpen
\bibfield  {author} {\bibinfo {author} {\bibfnamefont {K.}~\bibnamefont {Ptaszy\ifmmode~\acute{n}\else \'{n}\fi{}ski}},\ }\bibfield  {title} {\bibinfo {title} {Coherence-enhanced constancy of a quantum thermoelectric generator},\ }\href {https://doi.org/10.1103/PhysRevB.98.085425} {\bibfield  {journal} {\bibinfo  {journal} {Phys. Rev. B}\ }\textbf {\bibinfo {volume} {98}},\ \bibinfo {pages} {085425} (\bibinfo {year} {2018})}\BibitemShut {NoStop}%
\bibitem [{\citenamefont {Saryal}\ \emph {et~al.}(2019)\citenamefont {Saryal}, \citenamefont {Friedman}, \citenamefont {Segal},\ and\ \citenamefont {Agarwalla}}]{Saryal2019}%
\BibitemOpen
\bibfield  {author} {\bibinfo {author} {\bibfnamefont {S.}~\bibnamefont {Saryal}}, \bibinfo {author} {\bibfnamefont {H.~M.}\ \bibnamefont {Friedman}}, \bibinfo {author} {\bibfnamefont {D.}~\bibnamefont {Segal}},\ and\ \bibinfo {author} {\bibfnamefont {B.~K.}\ \bibnamefont {Agarwalla}},\ }\bibfield  {title} {\bibinfo {title} {Thermodynamic uncertainty relation in thermal transport},\ }\href@noop {} {\bibfield  {journal} {\bibinfo  {journal} {Phys. Rev. E}\ }\textbf {\bibinfo {volume} {100}},\ \bibinfo {pages} {042101} (\bibinfo {year} {2019})}\BibitemShut {NoStop}%
\bibitem [{\citenamefont {Saryal}\ \emph {et~al.}(2021)\citenamefont {Saryal}, \citenamefont {Gerry}, \citenamefont {Khait}, \citenamefont {Segal},\ and\ \citenamefont {Agarwalla}}]{Saryal2021}%
\BibitemOpen
\bibfield  {author} {\bibinfo {author} {\bibfnamefont {S.}~\bibnamefont {Saryal}}, \bibinfo {author} {\bibfnamefont {M.}~\bibnamefont {Gerry}}, \bibinfo {author} {\bibfnamefont {I.}~\bibnamefont {Khait}}, \bibinfo {author} {\bibfnamefont {D.}~\bibnamefont {Segal}},\ and\ \bibinfo {author} {\bibfnamefont {B.~K.}\ \bibnamefont {Agarwalla}},\ }\bibfield  {title} {\bibinfo {title} {Universal bounds on fluctuations in continuous thermal machines},\ }\href {https://doi.org/10.1103/PhysRevLett.127.190603} {\bibfield  {journal} {\bibinfo  {journal} {Phys. Rev. Lett.}\ }\textbf {\bibinfo {volume} {127}},\ \bibinfo {pages} {190603} (\bibinfo {year} {2021})}\BibitemShut {NoStop}%
\bibitem [{\citenamefont {Timpanaro}\ \emph {et~al.}(2025)\citenamefont {Timpanaro}, \citenamefont {Guarnieri},\ and\ \citenamefont {Landi}}]{Timpanaro2025}%
\BibitemOpen
\bibfield  {author} {\bibinfo {author} {\bibfnamefont {A.~M.}\ \bibnamefont {Timpanaro}}, \bibinfo {author} {\bibfnamefont {G.}~\bibnamefont {Guarnieri}},\ and\ \bibinfo {author} {\bibfnamefont {G.~T.}\ \bibnamefont {Landi}},\ }\bibfield  {title} {\bibinfo {title} {Quantum thermoelectric transmission functions with minimal current fluctuations},\ }\href {https://doi.org/10.1103/PhysRevB.111.014301} {\bibfield  {journal} {\bibinfo  {journal} {Phys. Rev. B}\ }\textbf {\bibinfo {volume} {111}},\ \bibinfo {pages} {014301} (\bibinfo {year} {2025})}\BibitemShut {NoStop}%
\bibitem [{\citenamefont {Brandner}\ \emph {et~al.}(2018)\citenamefont {Brandner}, \citenamefont {Hanazato},\ and\ \citenamefont {Saito}}]{Brandner2018}%
\BibitemOpen
\bibfield  {author} {\bibinfo {author} {\bibfnamefont {K.}~\bibnamefont {Brandner}}, \bibinfo {author} {\bibfnamefont {T.}~\bibnamefont {Hanazato}},\ and\ \bibinfo {author} {\bibfnamefont {K.}~\bibnamefont {Saito}},\ }\bibfield  {title} {\bibinfo {title} {Thermodynamic bounds on precision in ballistic multiterminal transport},\ }\href {https://doi.org/10.1103/PhysRevLett.120.090601} {\bibfield  {journal} {\bibinfo  {journal} {Phys. Rev. Lett.}\ }\textbf {\bibinfo {volume} {120}},\ \bibinfo {pages} {090601} (\bibinfo {year} {2018})}\BibitemShut {NoStop}%
\bibitem [{\citenamefont {Macieszczak}\ \emph {et~al.}(2018)\citenamefont {Macieszczak}, \citenamefont {Brandner},\ and\ \citenamefont {Garrahan}}]{macieszczak2018unified}%
\BibitemOpen
\bibfield  {author} {\bibinfo {author} {\bibfnamefont {K.}~\bibnamefont {Macieszczak}}, \bibinfo {author} {\bibfnamefont {K.}~\bibnamefont {Brandner}},\ and\ \bibinfo {author} {\bibfnamefont {J.~P.}\ \bibnamefont {Garrahan}},\ }\bibfield  {title} {\bibinfo {title} {Unified thermodynamic uncertainty relations in linear response},\ }\href@noop {} {\bibfield  {journal} {\bibinfo  {journal} {Phys. Rev Lett.}\ }\textbf {\bibinfo {volume} {121}},\ \bibinfo {pages} {130601} (\bibinfo {year} {2018})}\BibitemShut {NoStop}%
\bibitem [{\citenamefont {Saryal}\ \emph {et~al.}(2022)\citenamefont {Saryal}, \citenamefont {Mohanta},\ and\ \citenamefont {Agarwalla}}]{Saryal2022}%
\BibitemOpen
\bibfield  {author} {\bibinfo {author} {\bibfnamefont {S.}~\bibnamefont {Saryal}}, \bibinfo {author} {\bibfnamefont {S.}~\bibnamefont {Mohanta}},\ and\ \bibinfo {author} {\bibfnamefont {B.~K.}\ \bibnamefont {Agarwalla}},\ }\bibfield  {title} {\bibinfo {title} {Bounds on fluctuations for machines with broken time-reversal symmetry: A linear response study},\ }\href {https://doi.org/10.1103/PhysRevE.105.024129} {\bibfield  {journal} {\bibinfo  {journal} {Phys. Rev. E}\ }\textbf {\bibinfo {volume} {105}},\ \bibinfo {pages} {024129} (\bibinfo {year} {2022})}\BibitemShut {NoStop}%
\bibitem [{\citenamefont {Braggio}\ \emph {et~al.}(2011)\citenamefont {Braggio}, \citenamefont {Governale}, \citenamefont {Pala},\ and\ \citenamefont {K{\"o}nig}}]{braggio2011superconducting}%
\BibitemOpen
\bibfield  {author} {\bibinfo {author} {\bibfnamefont {A.}~\bibnamefont {Braggio}}, \bibinfo {author} {\bibfnamefont {M.}~\bibnamefont {Governale}}, \bibinfo {author} {\bibfnamefont {M.~G.}\ \bibnamefont {Pala}},\ and\ \bibinfo {author} {\bibfnamefont {J.}~\bibnamefont {K{\"o}nig}},\ }\bibfield  {title} {\bibinfo {title} {Superconducting proximity effect in interacting quantum dots revealed by shot noise},\ }\href@noop {} {\bibfield  {journal} {\bibinfo  {journal} {Solid State Communications}\ }\textbf {\bibinfo {volume} {151}},\ \bibinfo {pages} {155} (\bibinfo {year} {2011})}\BibitemShut {NoStop}%
\bibitem [{\citenamefont {Taddei}\ and\ \citenamefont {Fazio}(2023)}]{taddei2023thermodynamic}%
\BibitemOpen
\bibfield  {author} {\bibinfo {author} {\bibfnamefont {F.}~\bibnamefont {Taddei}}\ and\ \bibinfo {author} {\bibfnamefont {R.}~\bibnamefont {Fazio}},\ }\bibfield  {title} {\bibinfo {title} {Thermodynamic uncertainty relations for systems with broken time reversal symmetry: The case of superconducting hybrid systems},\ }\href {https://doi.org/10.1103/PhysRevB.108.115422} {\bibfield  {journal} {\bibinfo  {journal} {Phys. Rev. B}\ }\textbf {\bibinfo {volume} {108}},\ \bibinfo {pages} {115422} (\bibinfo {year} {2023})}\BibitemShut {NoStop}%
\bibitem [{\citenamefont {Kamijima}\ \emph {et~al.}(2021)\citenamefont {Kamijima}, \citenamefont {Otsubo}, \citenamefont {Ashida},\ and\ \citenamefont {Sagawa}}]{Kamijima2021}%
\BibitemOpen
\bibfield  {author} {\bibinfo {author} {\bibfnamefont {T.}~\bibnamefont {Kamijima}}, \bibinfo {author} {\bibfnamefont {S.}~\bibnamefont {Otsubo}}, \bibinfo {author} {\bibfnamefont {Y.}~\bibnamefont {Ashida}},\ and\ \bibinfo {author} {\bibfnamefont {T.}~\bibnamefont {Sagawa}},\ }\bibfield  {title} {\bibinfo {title} {Higher-order efficiency bound and its application to nonlinear nanothermoelectrics},\ }\href {https://doi.org/10.1103/PhysRevE.104.044115} {\bibfield  {journal} {\bibinfo  {journal} {Phys. Rev. E}\ }\textbf {\bibinfo {volume} {104}},\ \bibinfo {pages} {044115} (\bibinfo {year} {2021})}\BibitemShut {NoStop}%
\bibitem [{\citenamefont {Ohnmacht}\ \emph {et~al.}(2025)\citenamefont {Ohnmacht}, \citenamefont {Cuevas}, \citenamefont {Belzig}, \citenamefont {L\'opez}, \citenamefont {Lim},\ and\ \citenamefont {Kim}}]{Ohnmacht2024}%
\BibitemOpen
\bibfield  {author} {\bibinfo {author} {\bibfnamefont {D.~C.}\ \bibnamefont {Ohnmacht}}, \bibinfo {author} {\bibfnamefont {J.~C.}\ \bibnamefont {Cuevas}}, \bibinfo {author} {\bibfnamefont {W.}~\bibnamefont {Belzig}}, \bibinfo {author} {\bibfnamefont {R.}~\bibnamefont {L\'opez}}, \bibinfo {author} {\bibfnamefont {J.~S.}\ \bibnamefont {Lim}},\ and\ \bibinfo {author} {\bibfnamefont {K.~W.}\ \bibnamefont {Kim}},\ }\bibfield  {title} {\bibinfo {title} {Thermodynamic uncertainty relations in superconducting junctions},\ }\href {https://doi.org/10.1103/PhysRevResearch.7.L012075} {\bibfield  {journal} {\bibinfo  {journal} {Phys. Rev. Res.}\ }\textbf {\bibinfo {volume} {7}},\ \bibinfo {pages} {L012075} (\bibinfo {year} {2025})}\BibitemShut {NoStop}%
\bibitem [{\citenamefont {L\'opez}\ \emph {et~al.}(2023)\citenamefont {L\'opez}, \citenamefont {Lim},\ and\ \citenamefont {Kim}}]{Lopez2023}%
\BibitemOpen
\bibfield  {author} {\bibinfo {author} {\bibfnamefont {R.}~\bibnamefont {L\'opez}}, \bibinfo {author} {\bibfnamefont {J.~S.}\ \bibnamefont {Lim}},\ and\ \bibinfo {author} {\bibfnamefont {K.~W.}\ \bibnamefont {Kim}},\ }\bibfield  {title} {\bibinfo {title} {Optimal superconducting hybrid machine},\ }\href {https://doi.org/10.1103/PhysRevResearch.5.013038} {\bibfield  {journal} {\bibinfo  {journal} {Phys. Rev. Res.}\ }\textbf {\bibinfo {volume} {5}},\ \bibinfo {pages} {013038} (\bibinfo {year} {2023})}\BibitemShut {NoStop}%
\bibitem [{\citenamefont {Manzano}\ and\ \citenamefont {L{\ifmmode\acute{o}\else\'{o}\fi}pez}(2023)}]{Manzano2023Oct}%
\BibitemOpen
\bibfield  {author} {\bibinfo {author} {\bibfnamefont {G.}~\bibnamefont {Manzano}}\ and\ \bibinfo {author} {\bibfnamefont {R.}~\bibnamefont {L{\ifmmode\acute{o}\else\'{o}\fi}pez}},\ }\bibfield  {title} {\bibinfo {title} {{Quantum-enhanced performance in superconducting Andreev reflection engines}},\ }\href {https://doi.org/10.1103/PhysRevResearch.5.043041} {\bibfield  {journal} {\bibinfo  {journal} {Phys. Rev. Res.}\ }\textbf {\bibinfo {volume} {5}},\ \bibinfo {pages} {043041} (\bibinfo {year} {2023})}\BibitemShut {NoStop}%
\bibitem [{\citenamefont {Misaki}\ and\ \citenamefont {Nagaosa}(2021)}]{Misaki2021}%
\BibitemOpen
\bibfield  {author} {\bibinfo {author} {\bibfnamefont {K.}~\bibnamefont {Misaki}}\ and\ \bibinfo {author} {\bibfnamefont {N.}~\bibnamefont {Nagaosa}},\ }\bibfield  {title} {\bibinfo {title} {{Theory of the nonreciprocal Josephson effect}},\ }\href {https://doi.org/10.1103/PhysRevB.103.245302} {\bibfield  {journal} {\bibinfo  {journal} {Phys. Rev. B}\ }\textbf {\bibinfo {volume} {103}},\ \bibinfo {pages} {245302} (\bibinfo {year} {2021})}\BibitemShut {NoStop}%
\bibitem [{\citenamefont {Palmqvist}\ \emph {et~al.}(2024)\citenamefont {Palmqvist}, \citenamefont {Tesser},\ and\ \citenamefont {Splettstoesser}}]{Palmqvist2024}%
\BibitemOpen
\bibfield  {author} {\bibinfo {author} {\bibfnamefont {D.}~\bibnamefont {Palmqvist}}, \bibinfo {author} {\bibfnamefont {L.}~\bibnamefont {Tesser}},\ and\ \bibinfo {author} {\bibfnamefont {J.}~\bibnamefont {Splettstoesser}},\ }\href {https://arxiv.org/abs/2410.10793} {\bibinfo {title} {Kinetic uncertainty relations for quantum transport}} (\bibinfo {year} {2024}),\ \Eprint {https://arxiv.org/abs/2410.10793} {arXiv:2410.10793 [cond-mat.mes-hall]} \BibitemShut {NoStop}%
\bibitem [{\citenamefont {Wozny}\ and\ \citenamefont {Leijnse}(2025)}]{Wozny2025}%
\BibitemOpen
\bibfield  {author} {\bibinfo {author} {\bibfnamefont {S.}~\bibnamefont {Wozny}}\ and\ \bibinfo {author} {\bibfnamefont {M.}~\bibnamefont {Leijnse}},\ }\bibfield  {title} {\bibinfo {title} {Current noise in quantum dot thermoelectric engines},\ }\href {https://doi.org/10.1103/PhysRevB.111.075422} {\bibfield  {journal} {\bibinfo  {journal} {Phys. Rev. B}\ }\textbf {\bibinfo {volume} {111}},\ \bibinfo {pages} {075422} (\bibinfo {year} {2025})}\BibitemShut {NoStop}%
\bibitem [{\citenamefont {Zhang}\ and\ \citenamefont {Su}(2025)}]{Zhang2025}%
\BibitemOpen
\bibfield  {author} {\bibinfo {author} {\bibfnamefont {Y.}~\bibnamefont {Zhang}}\ and\ \bibinfo {author} {\bibfnamefont {S.}~\bibnamefont {Su}},\ }\href {https://arxiv.org/abs/2503.13851} {\bibinfo {title} {Thermodynamic uncertainty relations for three-terminal systems with broken time-reversal symmetry}} (\bibinfo {year} {2025}),\ \Eprint {https://arxiv.org/abs/2503.13851} {arXiv:2503.13851 [cond-mat.stat-mech]} \BibitemShut {NoStop}%
\bibitem [{\citenamefont {Brandner}\ and\ \citenamefont {Saito}(2025)}]{Brandner2025}%
\BibitemOpen
\bibfield  {author} {\bibinfo {author} {\bibfnamefont {K.}~\bibnamefont {Brandner}}\ and\ \bibinfo {author} {\bibfnamefont {K.}~\bibnamefont {Saito}},\ }\bibfield  {title} {\bibinfo {title} {Thermodynamic uncertainty relations for coherent transport},\ }\href {https://doi.org/10.1103/6nww-8wcp} {\bibfield  {journal} {\bibinfo  {journal} {Phys. Rev. Lett.}\ }\textbf {\bibinfo {volume} {135}},\ \bibinfo {pages} {046302} (\bibinfo {year} {2025})}\BibitemShut {NoStop}%
\bibitem [{\citenamefont {Potanina}\ \emph {et~al.}(2021)\citenamefont {Potanina}, \citenamefont {Flindt}, \citenamefont {Moskalets},\ and\ \citenamefont {Brandner}}]{Potanina2021}%
\BibitemOpen
\bibfield  {author} {\bibinfo {author} {\bibfnamefont {E.}~\bibnamefont {Potanina}}, \bibinfo {author} {\bibfnamefont {C.}~\bibnamefont {Flindt}}, \bibinfo {author} {\bibfnamefont {M.}~\bibnamefont {Moskalets}},\ and\ \bibinfo {author} {\bibfnamefont {K.}~\bibnamefont {Brandner}},\ }\bibfield  {title} {\bibinfo {title} {Thermodynamic bounds on coherent transport in periodically driven conductors},\ }\href {https://doi.org/10.1103/PhysRevX.11.021013} {\bibfield  {journal} {\bibinfo  {journal} {Phys. Rev. X}\ }\textbf {\bibinfo {volume} {11}},\ \bibinfo {pages} {021013} (\bibinfo {year} {2021})}\BibitemShut {NoStop}%
\bibitem [{\citenamefont {Lu}\ \emph {et~al.}(2022)\citenamefont {Lu}, \citenamefont {Wang}, \citenamefont {Peng}, \citenamefont {Wang}, \citenamefont {Jiang},\ and\ \citenamefont {Ren}}]{Lu2022}%
\BibitemOpen
\bibfield  {author} {\bibinfo {author} {\bibfnamefont {J.}~\bibnamefont {Lu}}, \bibinfo {author} {\bibfnamefont {Z.}~\bibnamefont {Wang}}, \bibinfo {author} {\bibfnamefont {J.}~\bibnamefont {Peng}}, \bibinfo {author} {\bibfnamefont {C.}~\bibnamefont {Wang}}, \bibinfo {author} {\bibfnamefont {J.-H.}\ \bibnamefont {Jiang}},\ and\ \bibinfo {author} {\bibfnamefont {J.}~\bibnamefont {Ren}},\ }\bibfield  {title} {\bibinfo {title} {Geometric thermodynamic uncertainty relation in a periodically driven thermoelectric heat engine},\ }\href {https://doi.org/10.1103/PhysRevB.105.115428} {\bibfield  {journal} {\bibinfo  {journal} {Phys. Rev. B}\ }\textbf {\bibinfo {volume} {105}},\ \bibinfo {pages} {115428} (\bibinfo {year} {2022})}\BibitemShut {NoStop}%
\bibitem [{\citenamefont {Mart{\'\i}n-Rodero}\ and\ \citenamefont {Levy~Yeyati}(2011)}]{martin2011josephson}%
\BibitemOpen
\bibfield  {author} {\bibinfo {author} {\bibfnamefont {A.}~\bibnamefont {Mart{\'\i}n-Rodero}}\ and\ \bibinfo {author} {\bibfnamefont {A.}~\bibnamefont {Levy~Yeyati}},\ }\bibfield  {title} {\bibinfo {title} {Josephson and andreev transport through quantum dots},\ }\href {https://doi.org/10.1080/00018732.2011.624266} {\bibfield  {journal} {\bibinfo  {journal} {Advances in Physics}\ }\textbf {\bibinfo {volume} {60}},\ \bibinfo {pages} {899} (\bibinfo {year} {2011})}\BibitemShut {NoStop}%
\bibitem [{\citenamefont {Rozhkov}\ and\ \citenamefont {Arovas}(2000)}]{rozhkov2000interacting}%
\BibitemOpen
\bibfield  {author} {\bibinfo {author} {\bibfnamefont {A.}~\bibnamefont {Rozhkov}}\ and\ \bibinfo {author} {\bibfnamefont {D.~P.}\ \bibnamefont {Arovas}},\ }\bibfield  {title} {\bibinfo {title} {Interacting-impurity josephson junction: Variational wave functions and slave-boson mean-field theory},\ }\href@noop {} {\bibfield  {journal} {\bibinfo  {journal} {Phys. Rev. B}\ }\textbf {\bibinfo {volume} {62}},\ \bibinfo {pages} {6687} (\bibinfo {year} {2000})}\BibitemShut {NoStop}%
\bibitem [{\citenamefont {Meng}\ \emph {et~al.}(2009)\citenamefont {Meng}, \citenamefont {Florens},\ and\ \citenamefont {Simon}}]{meng2009self}%
\BibitemOpen
\bibfield  {author} {\bibinfo {author} {\bibfnamefont {T.}~\bibnamefont {Meng}}, \bibinfo {author} {\bibfnamefont {S.}~\bibnamefont {Florens}},\ and\ \bibinfo {author} {\bibfnamefont {P.}~\bibnamefont {Simon}},\ }\bibfield  {title} {\bibinfo {title} {Self-consistent description of andreev bound states in josephson quantum dot devices},\ }\href@noop {} {\bibfield  {journal} {\bibinfo  {journal} {Phys. Rev. B}\ }\textbf {\bibinfo {volume} {79}},\ \bibinfo {pages} {224521} (\bibinfo {year} {2009})}\BibitemShut {NoStop}%
\bibitem [{\citenamefont {Eldridge}\ \emph {et~al.}(2010)\citenamefont {Eldridge}, \citenamefont {Pala}, \citenamefont {Governale},\ and\ \citenamefont {K{\"o}nig}}]{eldridge2010superconducting}%
\BibitemOpen
\bibfield  {author} {\bibinfo {author} {\bibfnamefont {J.}~\bibnamefont {Eldridge}}, \bibinfo {author} {\bibfnamefont {M.~G.}\ \bibnamefont {Pala}}, \bibinfo {author} {\bibfnamefont {M.}~\bibnamefont {Governale}},\ and\ \bibinfo {author} {\bibfnamefont {J.}~\bibnamefont {K{\"o}nig}},\ }\bibfield  {title} {\bibinfo {title} {Superconducting proximity effect in interacting double-dot systems},\ }\href@noop {} {\bibfield  {journal} {\bibinfo  {journal} {Phys. Rev. B}\ }\textbf {\bibinfo {volume} {82}},\ \bibinfo {pages} {184507} (\bibinfo {year} {2010})}\BibitemShut {NoStop}%
\bibitem [{\citenamefont {Pala}\ \emph {et~al.}(2007)\citenamefont {Pala}, \citenamefont {Governale},\ and\ \citenamefont {K{\"o}nig}}]{pala2007nonequilibrium}%
\BibitemOpen
\bibfield  {author} {\bibinfo {author} {\bibfnamefont {M.~G.}\ \bibnamefont {Pala}}, \bibinfo {author} {\bibfnamefont {M.}~\bibnamefont {Governale}},\ and\ \bibinfo {author} {\bibfnamefont {J.}~\bibnamefont {K{\"o}nig}},\ }\bibfield  {title} {\bibinfo {title} {Nonequilibrium josephson and andreev current through interacting quantum dots},\ }\href@noop {} {\bibfield  {journal} {\bibinfo  {journal} {New Journal of Physics}\ }\textbf {\bibinfo {volume} {9}},\ \bibinfo {pages} {278} (\bibinfo {year} {2007})}\BibitemShut {NoStop}%
\bibitem [{\citenamefont {de~Jong}\ and\ \citenamefont {Beenakker}(1994)}]{beenakker1994}%
\BibitemOpen
\bibfield  {author} {\bibinfo {author} {\bibfnamefont {M.~J.~M.}\ \bibnamefont {de~Jong}}\ and\ \bibinfo {author} {\bibfnamefont {C.~W.~J.}\ \bibnamefont {Beenakker}},\ }\bibfield  {title} {\bibinfo {title} {Doubled shot noise in disordered normal-metal--superconductor junctions},\ }\href {https://doi.org/10.1103/PhysRevB.49.16070} {\bibfield  {journal} {\bibinfo  {journal} {Phys. Rev. B}\ }\textbf {\bibinfo {volume} {49}},\ \bibinfo {pages} {16070} (\bibinfo {year} {1994})}\BibitemShut {NoStop}%
\bibitem [{\citenamefont {Anantram}\ and\ \citenamefont {Datta}(1996)}]{anantram1996current}%
\BibitemOpen
\bibfield  {author} {\bibinfo {author} {\bibfnamefont {M.~P.}\ \bibnamefont {Anantram}}\ and\ \bibinfo {author} {\bibfnamefont {S.}~\bibnamefont {Datta}},\ }\bibfield  {title} {\bibinfo {title} {Current fluctuations in mesoscopic systems with andreev scattering},\ }\href {https://doi.org/10.1103/PhysRevB.53.16390} {\bibfield  {journal} {\bibinfo  {journal} {Phys. Rev. B}\ }\textbf {\bibinfo {volume} {53}},\ \bibinfo {pages} {16390} (\bibinfo {year} {1996})}\BibitemShut {NoStop}%
\bibitem [{\citenamefont {Cao}\ \emph {et~al.}(2015)\citenamefont {Cao}, \citenamefont {Fang}, \citenamefont {Chen},\ and\ \citenamefont {Luo}}]{cao2015currents}%
\BibitemOpen
\bibfield  {author} {\bibinfo {author} {\bibfnamefont {Z.}~\bibnamefont {Cao}}, \bibinfo {author} {\bibfnamefont {T.-F.}\ \bibnamefont {Fang}}, \bibinfo {author} {\bibfnamefont {Q.}~\bibnamefont {Chen}},\ and\ \bibinfo {author} {\bibfnamefont {H.-G.}\ \bibnamefont {Luo}},\ }\bibfield  {title} {\bibinfo {title} {Currents and current correlations in a topological superconducting nanowire beam splitter},\ }\href {https://doi.org/10.1209/0295-5075/111/57002} {\bibfield  {journal} {\bibinfo  {journal} {EPL}\ }\textbf {\bibinfo {volume} {111}},\ \bibinfo {pages} {57002} (\bibinfo {year} {2015})}\BibitemShut {NoStop}%
\bibitem [{\citenamefont {Meir}\ and\ \citenamefont {Wingreen}(1992)}]{wingreen1992}%
\BibitemOpen
\bibfield  {author} {\bibinfo {author} {\bibfnamefont {Y.}~\bibnamefont {Meir}}\ and\ \bibinfo {author} {\bibfnamefont {N.~S.}\ \bibnamefont {Wingreen}},\ }\bibfield  {title} {\bibinfo {title} {Landauer formula for the current through an interacting electron region},\ }\href {https://doi.org/10.1103/PhysRevLett.68.2512} {\bibfield  {journal} {\bibinfo  {journal} {Phys. Rev. Lett.}\ }\textbf {\bibinfo {volume} {68}},\ \bibinfo {pages} {2512} (\bibinfo {year} {1992})}\BibitemShut {NoStop}%
\bibitem [{\citenamefont {Haug}\ \emph {et~al.}(2008)\citenamefont {Haug}, \citenamefont {Jauho} \emph {et~al.}}]{haug2008quantum}%
\BibitemOpen
\bibfield  {author} {\bibinfo {author} {\bibfnamefont {H.}~\bibnamefont {Haug}}, \bibinfo {author} {\bibfnamefont {A.-P.}\ \bibnamefont {Jauho}}, \emph {et~al.},\ }\href@noop {} {\emph {\bibinfo {title} {Quantum kinetics in transport and optics of semiconductors}}},\ Vol.~\bibinfo {volume} {2}\ (\bibinfo  {publisher} {Springer},\ \bibinfo {year} {2008})\BibitemShut {NoStop}%
\bibitem [{\citenamefont {Splettstoesser}\ \emph {et~al.}(2007)\citenamefont {Splettstoesser}, \citenamefont {Governale}, \citenamefont {K{\"o}nig}, \citenamefont {Taddei},\ and\ \citenamefont {Fazio}}]{splettstoesser2007pumping}%
\BibitemOpen
\bibfield  {author} {\bibinfo {author} {\bibfnamefont {J.}~\bibnamefont {Splettstoesser}}, \bibinfo {author} {\bibfnamefont {M.}~\bibnamefont {Governale}}, \bibinfo {author} {\bibfnamefont {J.}~\bibnamefont {K{\"o}nig}}, \bibinfo {author} {\bibfnamefont {F.}~\bibnamefont {Taddei}},\ and\ \bibinfo {author} {\bibfnamefont {R.}~\bibnamefont {Fazio}},\ }\bibfield  {title} {\bibinfo {title} {Pumping through a quantum dot in the proximity of a superconductor},\ }\href {https://doi.org/10.1103/PhysRevB.75.235302} {\bibfield  {journal} {\bibinfo  {journal} {Phys. Rev. B}\ }\textbf {\bibinfo {volume} {75}},\ \bibinfo {pages} {235302} (\bibinfo {year} {2007})}\BibitemShut {NoStop}%
\bibitem{SM}
See Supplemental Material at \url{https://journals.aps.org/prresearch/supplemental/...}
for additional details on the numerical methods and extended data,
which includes Refs.~\cite{wingreen1992, pala2007nonequilibrium,haug2008quantum,splettstoesser2007pumping,anantram1996current,cao2015currents,beenakker1994}.
\end{thebibliography}

\begin{thebibliography}{9}%
	\makeatletter
	\providecommand \@ifxundefined [1]{%
		\@ifx{#1\undefined}
	}%
	\providecommand \@ifnum [1]{%
		\ifnum #1\expandafter \@firstoftwo
		\else \expandafter \@secondoftwo
		\fi
	}%
	\providecommand \@ifx [1]{%
		\ifx #1\expandafter \@firstoftwo
		\else \expandafter \@secondoftwo
		\fi
	}%
	\providecommand \natexlab [1]{#1}%
	\providecommand \enquote  [1]{``#1''}%
	\providecommand \bibnamefont  [1]{#1}%
	\providecommand \bibfnamefont [1]{#1}%
	\providecommand \citenamefont [1]{#1}%
	\providecommand \href@noop [0]{\@secondoftwo}%
	\providecommand \href [0]{\begingroup \@sanitize@url \@href}%
	\providecommand \@href[1]{\@@startlink{#1}\@@href}%
	\providecommand \@@href[1]{\endgroup#1\@@endlink}%
	\providecommand \@sanitize@url [0]{\catcode `\\12\catcode `\$12\catcode
		`\&12\catcode `\#12\catcode `\^12\catcode `\_12\catcode `\%12\relax}%
	\providecommand \@@startlink[1]{}%
	\providecommand \@@endlink[0]{}%
	\providecommand \url  [0]{\begingroup\@sanitize@url \@url }%
	\providecommand \@url [1]{\endgroup\@href {#1}{\urlprefix }}%
	\providecommand \urlprefix  [0]{URL }%
	\providecommand \Eprint [0]{\href }%
	\providecommand \doibase [0]{http://dx.doi.org/}%
	\providecommand \selectlanguage [0]{\@gobble}%
	\providecommand \bibinfo  [0]{\@secondoftwo}%
	\providecommand \bibfield  [0]{\@secondoftwo}%
	\providecommand \translation [1]{[#1]}%
	\providecommand \BibitemOpen [0]{}%
	\providecommand \bibitemStop [0]{}%
	\providecommand \bibitemNoStop [0]{.\EOS\space}%
	\providecommand \EOS [0]{\spacefactor3000\relax}%
	\providecommand \BibitemShut  [1]{\csname bibitem#1\endcsname}%
	\let\auto@bib@innerbib\@empty
	\bibitem [{\citenamefont {Meir}\ and\ \citenamefont
		{Wingreen}(1992)}]{wingreen1992_SM}%
	\BibitemOpen
	\bibfield  {author} {\bibinfo {author} {\bibfnamefont {Y.}~\bibnamefont
			{Meir}}\ and\ \bibinfo {author} {\bibfnamefont {N.~S.}\ \bibnamefont
			{Wingreen}},\ }\href {\doibase 10.1103/PhysRevLett.68.2512} {\bibfield
		{journal} {\bibinfo  {journal} {Phys. Rev. Lett.}\ }\textbf {\bibinfo
			{volume} {68}},\ \bibinfo {pages} {2512} (\bibinfo {year}
		{1992})}\BibitemShut {NoStop}%
	\bibitem [{\citenamefont {Pala}\ \emph {et~al.}(2007)\citenamefont {Pala},
		\citenamefont {Governale},\ and\ \citenamefont
		{König}}]{pala2007nonequilibrium_SM}%
	\BibitemOpen
	\bibfield  {author} {\bibinfo {author} {\bibfnamefont {M.~G.}\ \bibnamefont
			{Pala}}, \bibinfo {author} {\bibfnamefont {M.}~\bibnamefont {Governale}}, \
		and\ \bibinfo {author} {\bibfnamefont {J.}~\bibnamefont {König}},\ }\href
	{\doibase 10.1088/1367-2630/9/8/278} {\bibfield  {journal} {\bibinfo
			{journal} {New J. Phys.}\ }\textbf {\bibinfo {volume} {9}},\ \bibinfo {pages}
		{278} (\bibinfo {year} {2007})}\BibitemShut {NoStop}%
	\bibitem [{foo()}]{footnote1_SM}%
	\BibitemOpen
	\href@noop {} {}\bibinfo {note} {With this definition of $\mathcal{S}$, the
		thermal noise equals $2k_B T\mathcal{G}$, where $\mathcal{G}$ is the linear
		conductance.}\BibitemShut {Stop}%
	\bibitem [{\citenamefont {Haug}\ and\ \citenamefont
		{Jauho}(2008)}]{haug2008quantum_SM}%
	\BibitemOpen
	\bibfield  {author} {\bibinfo {author} {\bibfnamefont {H.}~\bibnamefont
			{Haug}}\ and\ \bibinfo {author} {\bibfnamefont {A.-P.}\ \bibnamefont
			{Jauho}},\ }\href@noop {} {\emph {\bibinfo {title} {Quantum Kinetics in
				Transport and Optics of Semiconductors}}},\ \bibinfo {edition} {2nd}\ ed.\
	(\bibinfo  {publisher} {Springer},\ \bibinfo {address} {Berlin},\ \bibinfo
	{year} {2008})\BibitemShut {NoStop}%
	\bibitem [{\citenamefont {Splettstoesser}\ \emph {et~al.}(2007)\citenamefont
		{Splettstoesser}, \citenamefont {Governale}, \citenamefont {König},
		\citenamefont {Taddei},\ and\ \citenamefont
		{Fazio}}]{splettstoesser2007pumping_SM}%
	\BibitemOpen
	\bibfield  {author} {\bibinfo {author} {\bibfnamefont {J.}~\bibnamefont
			{Splettstoesser}}, \bibinfo {author} {\bibfnamefont {M.}~\bibnamefont
			{Governale}}, \bibinfo {author} {\bibfnamefont {J.}~\bibnamefont {König}},
		\bibinfo {author} {\bibfnamefont {F.}~\bibnamefont {Taddei}}, \ and\ \bibinfo
		{author} {\bibfnamefont {R.}~\bibnamefont {Fazio}},\ }\href {\doibase
		10.1103/PhysRevB.75.235302} {\bibfield  {journal} {\bibinfo  {journal} {Phys.
				Rev. B}\ }\textbf {\bibinfo {volume} {75}},\ \bibinfo {pages} {235302}
		(\bibinfo {year} {2007})}\BibitemShut {NoStop}%
	\bibitem [{\citenamefont {Cao}\ \emph {et~al.}(2015)\citenamefont {Cao},
		\citenamefont {Fang}, \citenamefont {Chen},\ and\ \citenamefont
		{Luo}}]{cao2015currents_SM}%
	\BibitemOpen
	\bibfield  {author} {\bibinfo {author} {\bibfnamefont {Z.}~\bibnamefont
			{Cao}}, \bibinfo {author} {\bibfnamefont {T.-F.}\ \bibnamefont {Fang}},
		\bibinfo {author} {\bibfnamefont {Q.}~\bibnamefont {Chen}}, \ and\ \bibinfo
		{author} {\bibfnamefont {H.-G.}\ \bibnamefont {Luo}},\ }\href {\doibase
		10.1209/0295-5075/111/57002} {\bibfield  {journal} {\bibinfo  {journal}
			{EPL}\ }\textbf {\bibinfo {volume} {111}},\ \bibinfo {pages} {57002}
		(\bibinfo {year} {2015})}\BibitemShut {NoStop}%
	\bibitem [{\citenamefont {Anantram}\ and\ \citenamefont
		{Datta}(1996)}]{anantram1996current_SM}%
	\BibitemOpen
	\bibfield  {author} {\bibinfo {author} {\bibfnamefont {M.~P.}\ \bibnamefont
			{Anantram}}\ and\ \bibinfo {author} {\bibfnamefont {S.}~\bibnamefont
			{Datta}},\ }\href {\doibase 10.1103/PhysRevB.53.16390} {\bibfield  {journal}
		{\bibinfo  {journal} {Phys. Rev. B}\ }\textbf {\bibinfo {volume} {53}},\
		\bibinfo {pages} {16390} (\bibinfo {year} {1996})}\BibitemShut {NoStop}%
	\bibitem [{\citenamefont {de~Jong}\ and\ \citenamefont
		{Beenakker}(1994)}]{beenakker1994_SM}%
	\BibitemOpen
	\bibfield  {author} {\bibinfo {author} {\bibfnamefont {M.~J.~M.}\
			\bibnamefont {de~Jong}}\ and\ \bibinfo {author} {\bibfnamefont {C.~W.~J.}\
			\bibnamefont {Beenakker}},\ }\href {\doibase 10.1103/PhysRevB.49.16070}
	{\bibfield  {journal} {\bibinfo  {journal} {Phys. Rev. B}\ }\textbf {\bibinfo
			{volume} {49}},\ \bibinfo {pages} {16070} (\bibinfo {year}
		{1994})}\BibitemShut {NoStop}%
	\bibitem [{\citenamefont {Brandner}\ and\ \citenamefont
		{Saito}(2025)}]{Brandner-2025_SM}%
	\BibitemOpen
	\bibfield  {author} {\bibinfo {author} {\bibfnamefont {K.}~\bibnamefont
			{Brandner}}\ and\ \bibinfo {author} {\bibfnamefont {K.}~\bibnamefont
			{Saito}},\ }\href {\doibase 10.1103/6nww-8wcp} {\bibfield  {journal}
		{\bibinfo  {journal} {Phys. Rev. Lett.}\ }\textbf {\bibinfo {volume} {135}},\
		\bibinfo {pages} {046302} (\bibinfo {year} {2025})}\BibitemShut {NoStop}%
\end{thebibliography}
\end{document}